\documentclass[letterpaper,british,english,a4paper,twoside,twocolumn,twocolumn,english,showpacs,preprintnumbers,nofootinbib,referee]{revtex4-1}
\usepackage[T1]{fontenc}
\usepackage[latin1]{inputenc}
\usepackage{fancyhdr}
\usepackage{verbatim}
\usepackage{amsmath}
\usepackage{graphicx}
\usepackage{amssymb}

\makeatletter

\newcommand{\noun}[1]{\textsc{#1}}

\@ifundefined{textcolor}{}
{%
 \definecolor{BLACK}{gray}{0}
 \definecolor{WHITE}{gray}{1}
 \definecolor{RED}{rgb}{1,0,0}
 \definecolor{GREEN}{rgb}{0,1,0}
 \definecolor{BLUE}{rgb}{0,0,1}
 \definecolor{CYAN}{cmyk}{1,0,0,0}
 \definecolor{MAGENTA}{cmyk}{0,1,0,0}
 \definecolor{YELLOW}{cmyk}{0,0,1,0}
 }

\usepackage{fancyhdr}

\makeatletter

\def\be{\begin{equation}}
\def\ee{\end{equation}}
\def\ba{\begin{eqnarray}}
\def\ea{\end{eqnarray}}
\newcommand{\lsim}{\mathrel{\rlap{\lower4pt\hbox{\hskip1pt$\sim$}}
    \raise1pt\hbox{$<$}}}
\newcommand{\gsim}{\mathrel{\rlap{\lower4pt\hbox{\hskip1pt$\sim$}}
    \raise1pt\hbox{$>$}}}
\@addtoreset{equation}{section} 
\renewcommand{\theequation}{\arabic{section}.\arabic{equation}}
\let\old@appendix=\appendix 
\def\appendix{%
   \old@appendix 
   \@addtoreset{equation}{section} 
   \def\theequation{\thesection\arabic{equation}}} %
\fancypagestyle{plain}{\fancyhead{}}

\makeatother

\makeatother

\usepackage{babel}

\begin{document}

\title{Separating expansion from contraction in spherically symmetric models
with a perfect-fluid:\\
Generalization of the Tolman-Oppenheimer-Volkoff condition and
application to models with a cosmological constant}

\author{José P. Mimoso }

\email{jpmimoso@cii.fc.ul.pt}

\affiliation{Departamento de F\'{\i}sica, Faculdade de Ciências da Universidade
de Lisboa, Centro de Astronomia e Astrof\'{\i}sica,\\
Universidade de Lisboa%
\thanks{Previously at the Centro de F\'{\i}sica Teórica e Computacional%
}, Av. Gama Pinto 2, 1649-003 Lisboa, Portugal}

\author{Morgan Le Delliou}

\thanks{Also at Centro de F\'{\i}sica Teórica e Computacional, Universidade
de Lisboa, Av. Gama Pinto 2, 1649-003 Lisboa, Portugal}

\email{Morgan.LeDelliou@uam.es, delliou@cii.fc.ul.pt}

\affiliation{Instituto de Física Teórica UAM/CSIC, Facultad de Ciencias, C-XI,
Universidad Autónoma de Madrid\\
 Cantoblanco, 28049 Madrid Spain}

\author{Filipe C. Mena}

\email{fmena@math.uminho.pt}

\affiliation{Centro de Matemática, Universidade do Minho, Campus de Gualtar, 4710-057
Braga, Portugal}


\preprint{IFT-UAM/CSIC-08-14}

\preprint{Version \today}

\date{Received 2 November 2009; Accepted 5 May 2010}
\begin{abstract}
We investigate spherically symmetric perfect-fluid spacetimes and
discuss the existence and stability of a dividing shell separating
expanding and collapsing regions. We perform a $3+1$ splitting and
obtain gauge invariant conditions relating the intrinsic spatial curvature
of the shells to the Misner-Sharp mass and to a function of the pressure
that we introduce and that generalizes the Tolman-Oppenheimer-Volkoff
equilibrium condition.\textbf{ }We find that surfaces fulfilling those
two conditions fit, locally, the requirements of a dividing shell
and we argue that cosmological initial conditions should allow its
global validity\textbf{. }We analyze the particular cases of the Lema\^{\i}tre-Tolman-Bondi
dust models with a cosmological constant as an example of a cold dark
matter model with a cosmological constant ($\Lambda$-CDM) and its
generalization to contain a central perfect-fluid core. These models
provide simple, but physically interesting illustrations of our results. 
\end{abstract}
\maketitle

\section{Introduction}

Models of structure formation generally assume that small local inhomogeneities
grow due to the gravitational instability, so that the overdensities
collapse and eventually form the \textquotedbl{}bound\textquotedbl{}
structures we observe in the present universe. Underlying this viewpoint
is the idea that the collapse of the overdensities departs from the
general expansion of the universe. This approach often relies on the
idea that a small overdensity can be approached as a closed patch
in an otherwise spatially flat Friedmann universe, and it claims that
Birkhoff's theorem justifies that, on the one hand, its evolution
is independent from the outside universe, and, on the other hand,
that the behavior of the outside Friedman universe is immune to the
collapse of the closed patch (see e.g. \citep{Peebles 1981,Padmanabhan 1993,Cattoen:2005dx}).
The collapse of overdensities has been extensively studied and most
works have been focused on the study of the formation both of small
structure (astrophysical objects) and of large-scale structure as
the outcome of the growth of small perturbations in a cosmological
context. The latter subject comprises the relativistic and Newtonian
analysis of the evolution of the fluctuations (see e.g. \citep{deSitter:1933,Hawking:1966qi,Mukhanov:1990me,Liddle:1993fq})
and the study of the subsequent amplification of the growing modes
into the nonlinear regime resorting to numerical methods (see e.g.
\citep{Bernardeau:2001qr,Mota:2004pa,LeDelliou:2005ig,Maor:2006rh}).
%
{}In the present work we consider spherically symmetric, inhomogeneous
universes with pressure and study the question of whether there exists
a dividing shell separating expanding and collapsing regions. 
Our goal bears a connection to the general problem of assessing the
influence of global physics into the local physics \citep{Ellis:2001cq,Faraoni:2007es}.
One aspect of this problem that has always attracted great interest
is the endeavor to explain the local inertial phenomena in a Machian
sense (see e.g. \citep{Sciama:1953zz,Dicke:1961ma}) and, in fact,
Brans-Dicke theory \citep{Brans:1961sx,Barrow:1994nx,Mimoso:1994wn,Iguchi:2004yh}
stems from this problem.

Another related aspect has been the study of the influence of cosmic
expansion on local systems. Einstein and Straus \citep{ES} were the
first to study this problem by constructing a global solution that
resulted from matching the spherically symmetric vacuum Schwarzschild
solution to an expanding dust Friedmann-Lemaître-Robertson-Walker
(FLRW) exterior across a hypersurface preserving the symmetry. Bonnor
has made several investigations along this line (see e.g. \citep{Bonnor1974}).
In particular, he copresented an exact solution representing a local
distribution of 
electrically counterpoised dust 
 embedded in an expanding universe with zero spatial curvature \citep{Bonnor 1996},
showing that the distribution participates in the expansion.%
{} Among the generalizations of this model are settings that keep the
spherical symmetry but generalize the interior source fields by considering,
for example, Vaidya (see \citep{Fayos} and references therein) or
Lemaître-Tolman-Bondi (LTB) spacetimes (see \citep{Krasinski,Krazinski 2001,Krasinski:2003yp,Hellaby:2005ut,Mena:2004ck}%
{}). On a different context, Herrera and co-workers \citep{Herrera92,DiPrisco94,Herrera}
have studied %
{} the \textquotedbl{}cracking\textquotedbl{} of compact objects in
astrophysics using small anisotropic perturbations around spherically
symmetric homogeneous fluids in equilibrium. The latter references
are concerned with the existence of a shell where there is a change
in the direction of the radial force acting on the particles of the
shells. Whenever this happens one has a cracking situation, a concept
introduced by Herrera in Ref. \citep{Herrera92}. The approach of
these works is somewhat complementary to ours because it is not the
full evolution that is depicted there, but rather the effect on particles
as a result from a departure from equilibrium. 

In this work we use a different approach from all the works described
above. On one hand, by making use of a single coordinate patch, we
do not have to handle the matching problem. On the other hand, our
approach is not perturbative. We adopt the formalism that has recently
been developed in a remarkable series of papers by Lasky and Lun using
generalized Painlevé-Gullstrand (GPG) coordinates \citep{Adler et al 05,LaskyLun06b,LaskyLun06+}.
We perform a $3+1$ splitting and obtain gauge invariant conditions
relating not only the intrinsic spatial curvature of the shells to
the Misner-Sharp mass%
\footnote{also referred to as ADM mass 
 when considering the mass of the whole spatial hypersurface.%
} \citep{MisnerSharp} but also a function of the pressure that we
introduce and that generalizes the Tolman-Oppenheimer-Volkoff (TOV)
equilibrium condition.

In particular, we consider that the existence of a spherical shell
separating an expanding outer region from an inner region collapsing
to the center of symmetry, depends essentially on two conditions.
The first condition establishes that there is no matter exchange across
the shell%
{}. The second condition establishes that the generalized TOV equation
is satisfied on that shell, and hence that this shell is in some sort
of equilibrium. The difference with respect to the original problem
where the TOV equation was introduced for the first time is twofold.
Our result does not rely on the assumption of a static equilibrium
of the spherical distribution of matter, and consequently does not
assume that all the internal spherical perfect-fluid spherical shells
are constrained to satisfy the TOV equation. In our case the generalized
TOV equation is just satisfied at the dividing shell. %
{}Besides, the generalized TOV function depends on the spatial 3-curvature
in a more general way than the original TOV equation. Furthermore,
we shall characterize the dividing shell with kinematic quantities
that provide a gauge invariant formulation of the problem.

In order to illustrate our results we will analyze some particular
cases. The simplest example is provided by the well-known Lema\^{\i}tre-Tolman-Bondi
dust models with a cosmological constant that can be seen as an example
of a $\Lambda$-CDM model. A preliminary presentation of this work
can be found in \citep{Delliou:2009dm}. As a second case we consider
generalizations of the previous model to contain a central perfect-fluid
core. These models provide simple, but physically interesting illustrations
of our results.%
{}

%
{}

An outline of the paper is as follows: Section II The GPG-formalism
of Lasky and Lun: $3+1$ splitting and gauge invariants kinematical
quantities. Section III Existence of a shell separating contraction
from expansion: general conditions. Section IV Particular examples:
Section IV A $\Lambda$-CDM model (LTB with a cosmological constant).
Section IV B Perfect-fluid core in a $\Lambda$-CDM model. Section
V Discussion of our results.

We shall use units such that $8\pi G=1=c$, and the following index
convention: Greek indices $\alpha,\beta,...=1,2,3$ while latin indices
$a,b,...=0,1,2,3$. 

\section{$3+1$ splitting and gauge invariants kinematical quantities}

In this section we set the basic equations that we shall subsequently
need. For comparison, we follow closely the formalism used by Lasky
and Lun \citep{LaskyLun06b}, while slightly generalizing their derivations
for the explicit presence of a cosmological constant $\Lambda$. 

\subsection{Metric and ADM splitting}

We adopt the GPG coordinates of Ref.~\citep{LaskyLun06b} and perform
an Arnowitt, Deser and Misner (ADM \citep{Arnowitt:1962hi}) 3+1 splitting
\citep{EllisElst98} in which the spherically symmetric line element
assumes a perfect-fluid timelike normalized flow $n_{a}:=-\alpha\nabla_{a}t=\left[-\alpha,0,0,0\right]$
($n_{a}n^{a}=-1$), defining with its lapse $N=\alpha$ and its radial
shift vector $N^{\mu}=\left(\beta,0,0\right)$, an evolution of the
spatially curved three-metric $^{3}g_{\mu\nu}=diag\left(\frac{1}{1+E},r^{2},r^{2}\sin^{2}\theta\right)$
with time ($d\Omega^{2}:=d\theta^{2}+\sin^{2}\theta d\phi^{2}$),
\begin{multline}
ds^{2}=-\alpha(t,r)^{2}dt^{2}+\frac{1}{1+E(t,r)}\left(\beta(t,r)dt+dr\right)^{2}\\
+r^{2}d\Omega^{2}.\label{eq:dsLaskyLun}\end{multline}
 The 3+1 approach uses the projection operators along and orthogonal
to the flow \begin{alignat}{2}
N_{\, b}^{a}:=-n^{a}n_{b} & ,~~~ & h^{ab}:= & g^{ab}+n^{a}n^{b}.\label{eq:ProjNh}\end{alignat}
 where $h^{ab}$ is the 3-metric on the surface $\Sigma$ normal to
the flow. Those projectors are also used for covariant derivatives:
Along the flow, the proper time derivative of any tensor $X_{\,\, cd}^{ab}$
is \begin{align}
\dot{X}_{\,\, cd}^{ab} & :=n^{e}X_{\,\, cd;e}^{ab},\end{align}
 and in the orthogonal 3-surface, each component is projected with
$h$\begin{align}
X_{\,\,\bar{c}\bar{d};\bar{e}}^{\bar{a}\bar{b}} & :=h_{f}^{a}h_{g}^{b}h_{c}^{i}h_{d}^{j}h_{e}^{k}X_{\,\, ij;k}^{fg}.\label{eq:Projectors}\end{align}
 Then the covariant derivative of the flow, from its projections,
is defined as\begin{multline}
n_{a;b}=N_{b}^{\, c}n_{a;c}+n_{\bar{a};\bar{b}}=-n_{b}\dot{n}_{a}+{\textstyle \frac{1}{3}}\Theta h_{ab}+\sigma_{ab}\\
+\omega_{ab},\end{multline}
 where the projection trace, the expansion of the flow, is $\Theta=n_{\,;\bar{a}}^{a}$,
the rate of shear $\sigma_{ab}$ is its symmetric trace-free part
and its skew-symmetric part is the vorticity $\omega_{ab}$.

For perfect-fluids we have the Raychaudhuri propagation equation\begin{multline}
\dot{\Theta}-\dot{n}_{\,;\bar{a}}^{a}=-\frac{1}{3}\Theta^{2}+\dot{n}^{a}\dot{n}_{a}-\sigma_{ab}\sigma^{ab}+\omega_{ab}\omega^{ab}\\
-\frac{\kappa}{2}\left(\rho+3P\right)+\Lambda.\end{multline}
 where $\kappa=8\pi$. 

The quantity $\Theta_{ab}:=\frac{1}{2}\mathcal{L}_{n}h_{ab}$, where
$\mathcal{L}_{n}$ is the Lie derivative along the vector field $n^{a}$,
is the so-called extrinsic curvature and is given by%
\footnote{\label{fn:Recall-that-for}Recall that for a scalar $\mathcal{L}_{n}=n^{a}\partial_{a}=\frac{1}{\alpha}\partial_{t}-\frac{\beta}{\alpha}\partial_{r}$;
\citep{LaskyLun06b} called it $K_{ab}$ but we preferred the Ellis
convention for the extrinsic curvature. The prime denotes partial
radial derivatives while the dot will denote from here on partial
time derivatives.%
} 
\begin{align}
\Theta^{ab}= & diag\left[0,-\frac{1+E}{\alpha}\aleph,-\frac{\beta}{\alpha r^{3}},-\frac{\beta}{\alpha r^{3}\sin^{2}\theta}\right],\\
\textrm{ with }\aleph= & \left[\beta^{\prime}+\frac{1}{2}\frac{\dot{E}-\beta E^{\prime}}{1+E}\right].\nonumber \end{align}
Its trace is the expansion scalar
%
\footnote{Note that we obtain a sign for $\Theta$ and $a$ different from that
of Ref.~\citep{LaskyLun06b}.%
} \begin{align}
\Theta= & -\frac{\left(\beta r^{2}\right)^{\prime}}{\alpha r^{2}}-\frac{1}{2}\frac{\mathcal{L}_{n}E}{1+E},\label{eq:Kexpr}\end{align}
which leads to the shear scalar
\begin{align}
a= & \frac{1}{3}\frac{r}{\alpha}\left(\frac{\beta}{r}\right)^{\prime}+\frac{1}{6}\frac{\mathcal{L}_{n}E}{1+E}.\label{eq:aDef}\end{align}


The 3-Ricci curvature tensor, which arises from fully projecting the
Riemann tensor in accordance with Eq.~(\ref{eq:Projectors}), is\begin{multline}
^{3}R_{\mu\nu}=diag\left[-\frac{E^{\prime}}{(1+E)r},-\frac{1}{2}E^{\prime}r-E,\right.\\
\left.\left(-\frac{1}{2}E^{\prime}r-E\right)\sin^{2}\theta\right].\end{multline}
 Then, the 3-Ricci trace and trace-free 3-Ricci tensor derive from
the 3-metric as\begin{align}
^{3}R= & -2\frac{\left(Er\right)^{\prime}}{r^{2}}\label{eq:3Rdef}\end{align}
 and \begin{align}
^{3}Q_{\mu\nu}:= & ^{3}R_{\mu\nu}-{\textstyle \frac{1}{3}}\:{}^{3}g_{\mu\nu}{}^{3}R\\
\Rightarrow{}^{3}Q_{\,\nu}^{\mu}= & \frac{1}{6}\frac{E^{\prime}r-2E}{r^{2}}P_{\,\nu}^{\mu}=q(t,r)P_{\,\nu}^{\mu}\label{Qab}\\
 & \Rightarrow q=\frac{r}{6}\left(\frac{E}{r^{2}}\right)^{\prime}.\label{eq:qDef}\end{align}
 where $P_{\,\nu}^{\mu}$ is $diag\left[-2,1,1\right]$.


The trace and trace-free Hessian of $\alpha$ are given by\begin{align}
\frac{1}{\alpha}D^{\mu}D_{\mu}\alpha= & \frac{\sqrt{1+E}}{\alpha r^{2}}\left(r^{2}\sqrt{1+E}\alpha^{\prime}\right)^{\prime}\label{eq:TrHessianAlpha}\end{align}
 and\begin{gather}
\frac{1}{\alpha}D_{\mu}D_{\nu}\alpha-\frac{1}{3\alpha}{}^{3}g_{\mu\nu}D^{c}D_{c}\alpha=\epsilon(t,r)P_{\mu\nu}\\
\textrm{ with }\epsilon=-\frac{r\sqrt{1+E}}{3\alpha}\left(\frac{\sqrt{1+E}}{r}\alpha^{\prime}\right)^{\prime},\label{eq:epsDef}\end{gather}
and where $D^{\mu}=h_{\nu}^{\mu}\nabla^{\nu}$ is the notation for
3-covariant derivative used in Ref.~\citep{Maartens97} and in Ref.~\citep{Ellis:2001cq}.

The Bianchi identity $T_{b;a}^{a}=0$ can be projected along $n^{b}$,
giving 
\begin{align}
n^{b}T_{b;a}^{a}= & -\mathcal{L}_{n}\rho-\left(\rho+P\right)\Theta=0.\end{align}
 while projections orthogonal to $n^{b}$ give the Euler equation\begin{align}
h_{a}^{\, b}T_{b;c}^{c}= & \left(\begin{array}{c}
\beta\\
1\\
0\\
0\end{array}\right)\left(P^{\prime}+\left(\rho+P\right)\frac{\alpha^{\prime}}{\alpha}\right)=0\label{hT}\\
\Rightarrow P^{\prime}= & -\left(\rho+P\right)\frac{\alpha^{\prime}}{\alpha}.\label{eq:Euler}\end{align}

\subsection{The Einstein field equations}

It is well known that the ADM approach separates the ten Einstein
field equations (EFE) into four constraints %
{} and six evolution equations. Spherical symmetry reduces them to 2+2
equations.


The Hamiltonian constraint reads, in the presence of a cosmological
constant,\begin{align}
^{3}R+{\textstyle \frac{2}{3}}\Theta^{2}-6a^{2}= & 16\pi\rho+2\Lambda,\label{eq:Hamiltonian}\end{align}
 the momentum constraint, restricted to the radial direction by symmetry,\begin{align}
\left(r^{3}a\right)^{\prime}= & -\frac{r^{3}}{3}\Theta^{\prime}\label{eq:momentum}\end{align}
 and the evolution equations can be reduced to%
\footnote{Note the sign differences in front of the Lie derivatives terms compared
with \citep{LaskyLun06b}; our results give a sign for $\dot{H}$
which is consistent with the Raychaudhuri equation restricted to the
FLRW case.%
}\begin{multline}
-2\mathcal{L}_{n}\Theta-\frac{1}{2}{}^{3}R-\Theta^{2}-9a^{2}+\frac{2}{\alpha}D^{a}D_{a}\alpha\\
=24\pi P-3\Lambda,\label{eq:evol1}\end{multline}
\vspace{-1cm}
\begin{align}
-\mathcal{L}_{n}a-a\Theta+\epsilon-q= & 0.\label{eq:evol2}\end{align}
Using Eqs.~(\ref{eq:Kexpr}) and (\ref{eq:aDef}) in Eq.~(\ref{eq:momentum}),
one can simplify the latter into\begin{align}
-\frac{\mathcal{L}_{n}E}{1+E}= & 2\frac{\beta}{\alpha^{2}}\alpha^{\prime}.\label{eq:momentum2}\end{align}
 Using the guidance that, from Eqs.~(\ref{eq:3Rdef}) and (\ref{eq:qDef}),
$^{3}R+12q$ eliminates derivatives in $E$, we can further simplify
the combination of Eqs.~{[}(\ref{eq:evol1}) + 6(\ref{eq:evol2}){]}$\times r^{2}/3$
with expressions from Eqs.~(\ref{eq:Kexpr}), (\ref{eq:aDef}), (\ref{eq:3Rdef}),
(\ref{eq:qDef}), and (\ref{eq:TrHessianAlpha}) as\begin{multline}
2r\left(1+E\right)\left(\ln\alpha\right)^{\prime}-8\pi Pr^{2}+\Lambda r^{2}+2r\mathcal{L}_{n}\left(\frac{\beta}{\alpha}\right)-\left(\frac{\beta}{\alpha}\right)^{2}\\
=-E.\label{eq:evol3}\end{multline}
 Substitution of Eq.~(\ref{eq:evol3}) into Eq.~(\ref{eq:Hamiltonian})$\times r^{2}/4$
yields, together with Eqs.~(\ref{eq:Kexpr}), (\ref{eq:aDef}), (\ref{eq:3Rdef}),
(\ref{eq:momentum2}), and $r/2\times$(\ref{eq:evol3}), a Poisson-like
equation that, integrated %
{}over $r$, defines a Misner-Sharp mass function \citep{MisnerSharp}\begin{multline}
M^{\prime}=4\pi\rho r^{2}\Rightarrow M=4\pi\int_{0}^{r}\rho x^{2}dx=r^{2}\left(1+E\right)\left(\ln\alpha\right)^{\prime}\\
-4\pi Pr^{3}+\frac{1}{3}\Lambda r^{3}+r^{2}\mathcal{L}_{n}\left(\frac{\beta}{\alpha}\right),\label{eq:Mdef}\end{multline}
 which with Euler's Eq.~(\ref{eq:Euler}) rewritten, for $P\ne-\rho$,
leads to the expression\begin{align}
\frac{M}{r^{2}}+4\pi Pr= & \mathcal{L}_{n}\left(\frac{\beta}{\alpha}\right)+\frac{1}{3}\Lambda r-\frac{1+E}{\rho+P}P^{\prime}.\label{eq:evol4}\end{align}
 The evolution Eq.~(\ref{eq:evol3}) can be recast to recognize the
definition of (\ref{eq:Mdef}):\begin{align}
E+2\frac{M}{r}+\frac{1}{3}\Lambda r^{2}= & \left(\frac{\beta}{\alpha}\right)^{2}.\label{eq:RadialEvol}\end{align}
%
{}With Euler's Eq.~(\ref{eq:Euler}) , the momentum Eq.~(\ref{eq:momentum2})
becomes\begin{align}
\mathcal{L}_{n}E= & 2\frac{\beta}{\alpha}\frac{1+E}{\rho+P}P^{\prime},\label{eq:LieE}\end{align}
%
{}while taking Eq.~(\ref{eq:RadialEvol})'s Lie derivative and using
(\ref{eq:LieE}) with $\mathcal{L}_{n}\frac{1}{r}=-\frac{\beta}{\alpha}\partial_{r}\frac{1}{r}=\frac{\beta}{\alpha}/r^{2}$,
then $\frac{\beta}{\alpha}\times$Eq.~(\ref{eq:evol4}) reads\begin{align}
\mathcal{L}_{n}M= & 4\pi Pr^{2}\frac{\beta}{\alpha}.\label{eq:LieM}\end{align}
%
{}Taking the positive (contracting) root of Eq.~(\ref{eq:RadialEvol}),
the evolution Eqs.~$\alpha\times$(\ref{eq:LieM}) and $\alpha\times$(\ref{eq:LieE})
for $M$ and $E$ can be written in terms of time derivatives where
we render explicit the Lie derivative (see footnote \ref{fn:Recall-that-for}):\begin{align}
\dot{M} & =\alpha\left(M^{\prime}+4\pi Pr^{2}\right)\sqrt{2\frac{M}{r}+\frac{1}{3}\Lambda r^{2}+E},\label{eq:Mdot}\end{align}
 \begin{multline}
\dot{E}=\alpha\left(E^{\prime}+2\frac{1+E}{\rho+P}P^{\prime}\right)\sqrt{2\frac{M}{r}+\frac{1}{3}\Lambda r^{2}+E}.\label{eq:Edot}\end{multline}
%
{}This system is then closed with a choice of an equation of state (%
{}EoS)%
{}.

\subsection{Generalized LTB}

Getting the metric (\ref{eq:dsLaskyLun}) into the generalized LTB
(GLTB) form, as in \citep{LaskyLun06b}, requires a coordinate transform
so that $\beta dt+dr\propto dR$. Taking $t(T)$ and $r(T,R)$, we
have then the condition \begin{equation}
\beta\partial_{T}t+\partial_{T}r=0,\label{eq:CondLaskyLTB}\end{equation}
 which becomes\begin{align}
\beta= & -\dot{r}.\label{eq:betaRdot}\end{align}
 Consequently, the line element (\ref{eq:dsLaskyLun}) can be rewritten
as\begin{multline}
ds^{2}=-\alpha(T,R)^{2}\left(\partial_{T}t\right)^{2}dT^{2}+\frac{\left(\partial_{R}r\right)^{2}}{1+E(T,R)}dR^{2}+r^{2}d\Omega^{2},\label{eq:dsLTB}\end{multline}
where $E(T,R)>-1$ and we can freely absorb the time function in the
new time by choosing $t=T$. Using now $\dot{\phantom{l}}$ and $\phantom{A}^{\prime}$
for $\partial_{T}$ and $\partial_{R}$, respectively, Eq.~(\ref{eq:RadialEvol})
now reads\begin{align}
\dot{r}^{2}= & \alpha^{2}\left(2\frac{M}{r}+\frac{1}{3}\Lambda r^{2}+E\right)\label{eq:RadEvolLTB}\end{align}
 and Eq.~(\ref{eq:Mdot}) rewrites, using Eq.~(\ref{eq:betaRdot}),\begin{align}
\dot{M}= & \beta4\pi Pr^{2}=4\pi Pr^{2}\alpha\sqrt{2\frac{M}{r}+\frac{1}{3}\Lambda r^{2}+E},\label{eq:MdotLTB}\end{align}
 while Eq.~(\ref{eq:Edot})$\times r^{\prime}$ rewrites\begin{align}
\dot{E}r^{\prime}= & 2\beta\frac{1+E}{\rho+P}P^{\prime}=2\frac{1+E}{\rho+P}P^{\prime}\alpha\sqrt{2\frac{M}{r}+\frac{1}{3}\Lambda r^{2}+E}\label{eq:EdotLTB}\end{align}
 and Euler's Eq.~(\ref{eq:Euler})$\times r^{\prime}$ is unchanged,\begin{align}
\frac{\alpha^{\prime}}{\alpha}= & -\frac{P^{\prime}}{\rho+P}.\label{eq:EulerLTB}\end{align}

\subsection{Remarks on $\Lambda$\label{sub:Remarks-on-Lambda}}

In all that precedes, the cosmological constant was kept explicit.
However, from the EFEs, one can include its effects in the total density
and pressure as that of a fluid with $\rho_{\Lambda}=-P_{\Lambda}=\frac{\Lambda}{\kappa}$.
We then obtain expressions identical to Lasky and Lun \citep{LaskyLun06b}.
%
{}It is interesting to note that the Misner-Sharp mass, in the explicit
$\Lambda$ formulation, is only referring to the initial, {}``$\Lambda$-less{}``
mixture, while encompassing the gravitational effects of the presence
of $\Lambda$. From Eq.~(\ref{eq:Mdef}) we can 
define the mass $M_{\mathrm{tot}}$ and pressure term $4\pi P_{\mathrm{tot}}r^{3}$
for the sum of the total perfect-fluid mixture plus $\Lambda$ by
taking Eq.~(\ref{eq:Mdef}) for a perfect-fluid and setting $\Lambda=0$.
We can also interpret the sum of the total mass and pressure terms
as the mass of an equivalent dust model $M_{\mathrm{ed}}$. We can
then 
integrate the mass of $\Lambda$ fluid and 
introduce the {}``Misner-Sharp mass'' \citep{MisnerSharp} pressure
term for the $\Lambda$ fluid: \begin{align}
M_{\mathrm{tot}}+4\pi P_{\mathrm{tot}}r^{3}= & r^{2}\left(1+E\right)\left(\ln\alpha\right)^{\prime}+r^{2}\mathcal{L}_{n}\left(\frac{\beta}{\alpha}\right)\equiv M_{\mathrm{ed}},\\
M_{\Lambda}= & \frac{4\pi}{3}r^{3}\rho_{\Lambda}=\frac{\Lambda}{6}r^{3},\\
4\pi P_{\Lambda}r^{3}= & -{\textstyle \frac{1}{2}}\Lambda r^{3}.\end{align}
 Thus we can rewrite the Misner-Sharp sum of the mass and pressure
term from its components from Eq.~(\ref{eq:Mdef}) :\begin{align}
M+4\pi Pr^{3}= & M_{\mathrm{tot}}+4\pi P_{\mathrm{tot}}r^{3}+{\textstyle \frac{1}{3}}\Lambda r^{3},\label{eq:MpPterm}\\
M_{\Lambda}+4\pi P_{\Lambda}r^{3}= & -\frac{1}{2}\Lambda r^{3}+\frac{\Lambda}{6}r^{3}=-\frac{1}{3}\Lambda r^{3},\label{eq:LambdaMpPterm}\end{align}
 so $M_{\mathrm{tot}}=M+M_{\Lambda}$ and $P_{\mathrm{tot}}=P+P_{\Lambda}$.
In Sec. \ref{sec:Geometrical-and-physical}, unless stated otherwise,
we will use $M$, $\rho$, and $P$ to refer to the \emph{total values}
of the corresponding quantities, while we will adopt the notation
$M_{pf}$, $\rho_{pf}$, and $P_{pf}$ to refer to the perfect-fluid
quantities. We also wish to remark that although the mass evolution
Eq.~(\ref{eq:LieM}) refers to the {}``$\Lambda$-less{}`` mixture
mass and pressure, this conservation equation holds for each component
of a mixture of noncoupled fluids. We thus have for independent fluids

\begin{align}
M= & \sum_{\mathrm{fluid}\, i}M_{i},\\
P= & \sum_{\mathrm{fluid}\, i}P_{i},\\
\mathcal{L}_{n}M_{i}= & 4\pi P_{i}r^{2}\frac{\beta}{\alpha}=\pm4\pi P_{i}r^{2}\sqrt{2\frac{M}{r}+E}.\end{align}

\section{Geometrical and physical conditions for the existence of a dividing
shell\label{sec:Geometrical-and-physical}}

%
{} In our spherical symmetric approach, we are looking for shells dividing
expansion at all time from regions of mixed behavior involving periods
of collapse.

This leads to an investigation of the conditions for the dynamical
separation of sections of matter trapped inside a dividing surface
(physical condition)%
{}. We will see that this approach is distinct from a purely kinematic
separation of contraction from expansion (geometrical condition) and
will express the physical condition using kinematic quantities.%
{}

\subsection{Misner-Sharp mass conservation}

In the previous section we have seen how the Misner-Sharp mass%
{} is evolving with the flow under Eq.~(\ref{eq:LieM}). We can thus
define a surface for which this mass is conserved with respect to
the flow:

\begin{align}
\forall t,\,\mathcal{L}_{n}M(t,r_{\star}(t))= & 0\nonumber \\
\Leftrightarrow\forall t,\, E=-2\frac{M}{r_{\star}},\textrm{ or } & P_{\star}=0\textrm{ or }r_{\star}=0,\label{eq:cases}\end{align}
%
{}%
{}While the second case
, $P=0$, defines a dustlike layer in the perfect-fluid mix, and %
{}the third case, $r=0$, is trivial, %
{}we shall concentrate on the first case, $E=-2\frac{M}{r}$. In this
case, %
{} from Eq.~(\ref{eq:LieE}) we get \begin{align}
\mathcal{L}_{n}E= & \pm2\sqrt{2\frac{M}{r}+E}\frac{1+E}{\rho+P}P^{\prime}=0,\label{eq:LieEcases}\end{align}
 so the shell is characterized by fixed curvature and Misner-Sharp
mass. This implies that if a prescribed initial $P$ and $\rho$ distribution
is given such that there exists a shell where \begin{align}
E_{\star} & =-2\frac{M_{\star}}{r_{\star}},\label{eq:LimitShell}\end{align}
 then this shell can locally separate inner and outer regions that
can be expanding and contracting differently. We call the separating
shell a {}``limit shell,'' and denote it with $\star$%
{}. %
{}In GPG coordinates %
{}the above condition is equivalent to $\left.\frac{\beta}{\alpha}\right|_{\star}=0$,
or to %
{} $\beta_{\star}=0$. %
{} We can then use it to compute%
{} \begin{align}
\dot{r}_{\star}= & -\frac{2M}{E}\alpha\left[\frac{\mathcal{L}_{n}M}{M}-\frac{\mathcal{L}_{n}E}{E}\right]_{\star}=0,\label{eq:rdot}\\
\ddot{r}_{\star}= & -\frac{2M}{E}\alpha^{2}\left[\frac{\mathcal{L}_{n}^{2}M}{M}-\frac{\mathcal{L}_{n}^{2}E}{E}\right]_{\star},\label{eq:rddot}\end{align}
and\begin{align}
\mathcal{L}_{n}r=-\frac{\beta}{\alpha} & \Rightarrow\mathcal{L}_{n}r_{\star}=0,\label{eq:LieR}\end{align}
%
{}so the limit shell appears as a {}``turnaround''%
\footnote{See discussion in \citep{Peebles  1981} Sec. 19, p. 77.%
} shell,%
{} in terms of areal radius.

However, these conditions are coordinate dependent and give limited
insight as to how they would express for different observers. This
calls for a definition using gauge invariant quantities.

\subsection{Expansion and shear}

Newtonian structure formation in spherical symmetry provides a natural
limiting shell that is a locus separating at a given time expansion
from collapse: the turnaround %
{} radius (see e.g.\citep{LeDH03}). The definition of that locus is
given by the vanishing of the expansion with respect to the flow.
Nevertheless, this is not necessarily the case resulting from condition
\ref{eq:cases}%
{}. Let us first start from the previous mass flow definition and examine
the corresponding expansion. %
{}

In GPG coordinates \citep{LaskyLun06b}, defining the flow by the
shift/lapse vector, we can compute the expansion (%
{}the trace of the symmetric part of the projected covariant derivative
of the flow vector), using Eqs.~(\ref{eq:momentum2}) and (\ref{eq:Kexpr}):\begin{align}
\Theta= & -\left(\frac{\beta}{\alpha}\right)^{\prime}-2\frac{\beta}{\alpha}\frac{1}{r}\label{eq:ThetaMetric1}\end{align}
%
{}At $r_{*}$ (for%
{} $\frac{\beta}{\alpha}=0$), we have nonzero expansion given by\begin{align}
\Theta_{\star}= & -\left(\frac{\beta}{\alpha}\right)_{\star}^{\prime}.\label{eq:ThetaStar}\end{align}
%
{}

The shear can also be expressed here from Eqs.~(\ref{eq:aDef}) and
(\ref{eq:momentum2}) as\begin{align}
a= & \frac{1}{3}\left[\left(\frac{\beta}{\alpha}\right)^{\prime}-\frac{\beta}{\alpha}\frac{1}{r}\right],\label{eq:aMetric1}\end{align}
and we can then relate shear and expansion as {[}using Eq.~\ref{eq:LieR}{]}
\begin{align}
r\left(\frac{\Theta}{3}+a\right)= & -\frac{\beta}{\alpha}=\mathcal{L}_{n}r,\label{eq:ExpLieR}\end{align}
%
{}so on the limit shell, \begin{align}
\Theta_{\star}+3a_{\star}= & 0\Leftrightarrow\left(\mathcal{L}_{n}r\right)_{\star}=0.\label{eq:LimitDef}\end{align}



\subsubsection{Generalizing $\mathrm{TOV}$}

The TOV equation, following \citep{LaskyLun06b}, emerges from Eq.~(\ref{eq:evol4})
in the static case.

%
{}

We now generalize the TOV equation by defining a functional $\mathrm{gTOV}$
from Eq.~(\ref{eq:evol4}) as\begin{equation}
\mathrm{gTOV}=\left[\frac{1+E}{\rho_{pf}+P_{pf}}P_{pf}^{\prime}+4\pi P_{pf}r+\frac{M_{pf}}{r^{2}}-\frac{1}{3}\Lambda r\right]\end{equation}
 Using Eqs.~(\ref{eq:MpPterm}) and (\ref{eq:LambdaMpPterm}) we
also have \begin{equation}
\mathrm{gTOV}=\left[\frac{1+E}{\rho+P}P^{\prime}+4\pi Pr+\frac{M}{r^{2}}\right].\label{eq:TOVdef}\end{equation}
%
{}The definitions (\ref{eq:ExpLieR}), (\ref{eq:evol4}), and (\ref{eq:TOVdef})
combine to yield\begin{gather}
\mathrm{gTOV}=-r\left[\mathcal{L}_{n}\left(\frac{\Theta}{3}+a\right)-\left(\frac{\Theta}{3}+a\right)^{2}\right]\\
=-\mathcal{L}_{n}^{2}r.\label{eq:Lie2r}\end{gather}
So, $\mathrm{gTOV}$ is equal to the radial acceleration or, more
generally, to the Lie derivative of $\beta/\alpha$, and hence Eq.(\ref{eq:Lie2r})
is the version in the GPG formalism of the classical Euler's equation
of continuum mechanics. We also see that this $\mathrm{gTOV}$ acceleration
relates to the force envisaged in the works of Herrera and collaborators
\citep{Herrera92,DiPrisco94,Herrera} multiplied by $(1+E)/\left(\rho+p\right)$,
i.e., by $(1-2M/r_{\star})/(\rho+p)$ at $r=r_{\star}$. We can then
obtain local %
{} conditions that yield the TOV equation on the limit shell when\begin{align}
\mathrm{gTOV}_{\star}=0 & \Leftrightarrow & \mathcal{L}_{n}^{2}r= & 0\nonumber \\
 & \Leftrightarrow & \mathcal{L}_{n}\left(\frac{\Theta}{3}+a\right)_{\star}= & 0.\label{eq:Lie2rStar}\end{align}

%
{}We can further express $\mathrm{gTOV}$ in a form that reminds us
of the FLRW Raychaudhuri equation by using $\left\langle \rho\right\rangle \equiv M/(4\pi r^{3}/3)$,
i.e.%
{}\begin{align}
\mathrm{gTOV}= & \frac{1+E}{\rho+P}P^{\prime}+\frac{4\pi}{3}r\left(\left\langle \rho\right\rangle +3P\right),\end{align}
%
{}and for FLRW %
{} it reduces to%
{}\begin{align}
\mathrm{gTOV}_{\mathrm{FL}}= & \frac{4\pi}{3}r\left(\rho+3P\right)=-\ddot{r}.\end{align}

\subsubsection{Dynamics of the limit shell}

We have seen that we could define the limit shell by only setting
$E_{\star}=-2M_{\star}/r_{\star}$ (so $\beta_{\star}=0$), so that
$\Theta_{\star}=3a_{\star}$. Now, using Eqs.~(\ref{eq:RadialEvol}),
(\ref{eq:Mdot}), (\ref{eq:Edot}), and (\ref{eq:TOVdef}), we find\begin{align}
\left(\frac{\beta}{\alpha}\right)^{\centerdot}= & \beta\left(\frac{\beta}{\alpha}\right)^{\prime}+\alpha\mathrm{gTOV}\\
\Rightarrow\dot{\beta}= & \beta\left(\beta^{\prime}-\beta\frac{\alpha^{\prime}}{\alpha}+\frac{\dot{\alpha}}{\alpha}\right)+\alpha^{2}\mathrm{gTOV},\end{align}
so on the limit shell, we have\begin{align}
\left(\frac{\beta}{\alpha}\right)_{\star}^{\centerdot}= & \alpha\mathrm{gTOV}_{\star}\\
\Rightarrow\dot{\beta}_{\star}= & \alpha^{2}\mathrm{gTOV}_{\star}.\end{align}
 Recall that, in the LTB frame, $\beta=-\dot{r}$, so this tells us\begin{align}
\ddot{r}_{\mathrm{LTB},\star}= & -\alpha^{2}\mathrm{gTOV}_{\star},\end{align}
 and thus when $\mathrm{gTOV}_{\star}=0$ that shell has no acceleration
and is therefore really static, as expressed in the original TOV equation.
For completeness, we can reexpress Eq.~(\ref{eq:LieR}) with Eqs.~(\ref{eq:LieM},\ref{eq:LieE},\ref{eq:TOVdef})
in GPG coordinates:%
{} \begin{align}
\ddot{r}_{\mathrm{GPG},\star} & =-\frac{2M}{E}\alpha^{2}\left[\frac{\mathcal{L}_{n}^{2}M}{M}-\frac{\mathcal{L}_{n}^{2}E}{E}\right]_{\star}\nonumber \\
 & =-\alpha^{2}\left[\mathrm{gTOV}_{\star}-r_{\star}^{2}\frac{\mathrm{gTOV}_{\star}^{2}}{M_{\star}}\right].\end{align}
From Eqs.~(\ref{eq:ThetaMetric1}) we derive upon integration\begin{equation}
\left(\frac{\beta}{\alpha}\right)=-r\left(a+\frac{\Theta}{3}\right)=\left(\frac{\beta}{\alpha}\right)_{r_{0}}\left(\frac{r_{0}}{r}\right)^{2}-\frac{1}{r^{2}}\int_{r_{0}}^{r}\,\Theta\, r^{2}{\rm d}r\label{eq:NonLocalTheta}\end{equation}
 where $(\beta/\alpha)_{r_{0}}$ is a function only of $t$, which
arises as the {}``constant'' of the integration performed with respect
to $r$, and which sets the value of $\beta/\alpha$ at $r=r_{0}$.
Using Eq. (\ref{eq:momentum}) integrated directly, or Eq.~(\ref{eq:NonLocalTheta})
and (\ref{eq:ExpLieR}), the latter result translates into\begin{align}
\left(a+\frac{\Theta}{3}\right)= & \left(\frac{r_{0}}{r}\right)^{3}\,\left(a_{r_{0}}+\frac{\Theta_{r_{0}}}{3}\right)+\frac{1}{r^{3}}\,\int_{r_{0}}^{r}\,\Theta\, r^{2}{\rm d}r\,,\end{align}
 which is its gauge invariant expression.

From Eq. (\ref{eq:NonLocalTheta}) we obtain\begin{multline}
\mathcal{L}_{n}\left(\frac{\beta}{\alpha}\right)=\left(\frac{\beta}{\alpha}\right)\,\left[\frac{2}{r^{3}}\left(r_{0}^{2}\left(\frac{\beta}{\alpha}\right)_{r_{0}}-\int_{r_{0}}^{r}\,\Theta\, r^{2}{\rm d}r\right)+\Theta\right]\\
+\frac{1}{\alpha r^{2}}\left[r_{0}^{2}\partial_{t}\left(\frac{\beta}{\alpha}\right)_{r_{0}}-\int_{r_{0}}^{r}\,\dot{\Theta}\, r^{2}{\rm d}r\right]=\mathrm{gTOV}\label{eq:NonLocalThetaDot}\end{multline}
 This is the general equation that corresponds indeed to Eq. (21)
of Di Prisco \emph{et al.} \citep{DiPrisco94}, and it confirms their
claim of a nonlocality of the radial acceleration. From Eq.~(\ref{eq:NonLocalTheta})
we realize that this nonlocality also characterizes the radial expansion,
as one should expect, and we further remark that a similar nonlocality
is already present in the energy condition defining $r_{\star}$ Eqs.~(\ref{eq:LimitShell})
and (\ref{eq:LimitDef}) and in our $\mathrm{gTOV}$ condition Eqs.~(\ref{eq:TOVdef})
and (\ref{eq:Lie2rStar}), since both implicate $M$, which is an
integral between $0$ and $r_{\star}$. 

As $\beta/\alpha=r(\Theta/3+a)$ vanishes at $r=r_{\star}$, from
Eq.~(\ref{eq:NonLocalTheta}) one deduces that\begin{equation}
\left(\frac{\beta}{\alpha}\right)_{r_{0}}r_{0}^{2}=\int_{r_{0}}^{r_{\star}}\,\Theta\, r^{2}{\rm d}r\;,\end{equation}
so that the integral on the right-hand side vanishes if the initial
parameter $r_{0}^{2}\left({\beta}/{\alpha}\right)_{r_{0}}$ vanishes
at some interior value $r_{0}<r_{\star}$. This implies that $\Theta$
must vanish at some intermediate value $r_{0}<r<r_{\star}$, since
it has to change signs within the interval of integration. %
{}

Likewise, when $\mathrm{gTOV}=0$, i.e. $\mathcal{L}_{n}\left(\beta/\alpha\right)$,
vanishes at some $r$, we derive from Eq.~(\ref{eq:NonLocalThetaDot})
that \begin{multline}
\left(\frac{\beta}{\alpha}\right)\,\left[\frac{2}{r^{3}}\left(r_{0}^{2}\left(\frac{\beta}{\alpha}\right)_{r_{0}}-\int_{r_{0}}^{r}\,\Theta\, r^{2}{\rm d}r\right)+\Theta\right]=\\
-\frac{1}{\alpha r^{2}}\left[r_{0}^{2}\partial_{t}\left(\frac{\beta}{\alpha}\right)_{r_{0}}-\int_{r_{0}}^{r}\,\dot{\Theta}\, r^{2}{\rm d}r\right]\;.\label{C}\end{multline}
So, at $r=r_{\star}$, the latter Eq.(\ref{C}) reduces to \begin{equation}
r_{0}^{2}\partial_{t}\left(\frac{\beta}{\alpha}\right)_{r_{0}}=\int_{r_{0}}^{r_{\star}}\,\dot{\Theta}\, r^{2}{\rm d}r\label{D}\end{equation}
 and we conclude that the integral on the right-hand side vanishes
if the term $r_{0}^{2}\partial_{t}(\beta/\alpha)_{r_{0}}$ vanishes
at $r_{0}$. This result shows that the vanishing of the time derivative
of the expansion thus occurs at least at one intermediate value between
$r_{0}$ and $r$. In the case when $\partial_{t}(\beta/\alpha)_{r_{0}}r_{0}^{2}=0$
at the center, we recover the result of Di Prisco \emph{et al.} \citep{DiPrisco94},
establishing the vanishing of the radial aceleration, i.e.  $\dot{\Theta}=0$,
at some $0<r<r_{\star}$.

\subsubsection{Raychaudhuri expansion evolution}

From Eqs.~(\ref{eq:Hamiltonian}) and (\ref{eq:evol1}), with $\Lambda$
included as a fluid component, we have%
{} in the GPG frame,\begin{align}
-2\mathcal{L}_{n}\Theta-\frac{2}{3}\Theta^{2}-12a^{2}+\frac{2}{\alpha}D^{k}D_{k}\alpha= & 8\pi\left(\rho+3P\right),\label{eq:Raychaudhuri}\end{align}
 and on the limit shell, that reads\begin{align}
-\frac{2}{\alpha}\dot{\Theta}_{\star}-2\Theta_{\star}^{2}+\frac{2}{\alpha}D^{k}D_{k}\alpha_{\star}= & 8\pi\left(\rho+3P\right),\end{align}
showing that this shell can still be dynamic. Using the Euler Eq.~(\ref{eq:Euler}),
the Hessian (\ref{eq:TrHessianAlpha}) gives%
{}\begin{align}
\frac{2}{\alpha}D^{\gamma}D_{\gamma}\alpha= & \frac{1+E}{\rho+P}P^{\prime}\left[\frac{E^{\prime}}{1+E}-\frac{2\left(\alpha r^{2}\right)^{\prime}}{\alpha r^{2}}\right]\nonumber \\
 & -2\left(\frac{1+E}{\rho+P}P^{\prime}\right)^{\prime}.\end{align}
 Thus Eq.~(\ref{eq:Raychaudhuri}) reads\begin{multline}
-\mathcal{L}_{n}\Theta-\Theta^{2}-\frac{2}{r}\frac{\beta}{\alpha}\left[2\Theta+\frac{3}{r}\frac{\beta}{\alpha}\right]=\\
4\pi\left(\rho+3P\right)-\frac{P^{\prime}}{2\left(\rho+P\right)}E^{\prime}+\left(\frac{1+E}{\rho+P}P^{\prime}\right)^{\prime}\\
+\left(\frac{2}{r}-\frac{P^{\prime}}{\rho+P}\right)\frac{1+E}{\rho+P}P^{\prime}.\label{eq:RaychaudhuriExplicit}\end{multline}
 Here, we can recognize the first term of $\mathrm{TOV}$. On the
limit shell the above equation reads\begin{align}
-\frac{1}{\alpha}\dot{\Theta}_{\star}-\Theta_{\star}^{2}= & 4\pi\left(\rho+3P\right)-\frac{P^{\prime}}{2\left(\rho+P\right)}E^{\prime}\nonumber \\
 & +\left(\frac{1+E}{\rho+P}P^{\prime}\right)^{\prime}+\left(\frac{2}{r}-\frac{P^{\prime}}{\rho+P}\right)\frac{1+E}{\rho+P}P^{\prime},\end{align}
%
{}and we re%
{}cast the Raychaudhuri equation for the FLRW case%
{}\begin{align}
-\mathcal{L}_{n}\Theta-\frac{\Theta^{2}}{3}= & 4\pi\left(\rho+3P\right)\\
= & -3\dot{H}-3H^{2}.\end{align}

\subsubsection{Remarks on null expansion limit shells\label{sub:Remarks-on-null}}

%
{}We now explore the consequences of having, in addition to (\ref{eq:LimitDef}),
the condition %
{} $\Theta_{\star}=0$%
{}for the limit shell. In this case, the shear must also vanish on the
shell%
{} and\begin{align}
\left(\frac{\beta}{\alpha}\right)_{\star}^{\prime}= & 0,\label{eq:shearFreeCond}\end{align}
which constrains the gradient of the generalized velocity field $\beta/\alpha$.%
{}

In addition, and most importantly, the Raychaudhuri Eq.~(\ref{eq:RaychaudhuriExplicit})
shows that an initially expansion-free dividing shell is not likely
to remain so, and will drift radially%
{}. If we impose %
{}the vanishing of $\mathcal{L}_{n}\Theta$ in Eq.~(\ref{eq:Raychaudhuri}),
we derive \begin{align}
\frac{1}{\alpha_{\star}}D^{k}D_{k}\alpha_{\star}= & 4\pi\left(\rho+3P\right)_{\star}\;,\end{align}
which then translates into a thermodynamic condition on the second-order
derivative of $P$, which should induce a very specific and \emph{ad
hoc} local equation of state of the perfect-fluid, namely\begin{align}
\left(\frac{1+E}{\rho+P}P^{\prime}\right)_{\star}^{\prime}= & -4\pi\left(\rho+3P\right)_{\star}+\frac{P_{\star}^{\prime}}{2\left(\rho+P\right)_{\star}}E_{\star}^{\prime}\nonumber \\
 & -\left(\frac{2}{r}-\frac{P^{\prime}}{\rho+P}\right)_{\star}\frac{1+E_{\star}}{\rho_{\star}+P_{\star}}P_{\star}^{\prime}.\label{eq:Hessean2staticity}\end{align}
%
{}We conclude that the case of a static, expansion-free, limit shell
is very restrictive: for example, in the simplest case, discussed
below, of an inhomogeneous $\Lambda$-CDM model, Eq.~(\ref{eq:Hessean2staticity})
induces a restrictive equation of state $P=-\rho/3$ on the shell,
which is verified neither by the dust component nor by the $\Lambda$
fluid, whereas the limit shell in this case derives from a staticity
condition (see Sec.~\ref{sub:Overdensity-in-a}).%
{}

\section{Applications to simple models}


We now will illustrate the behavior according to the limit shell of
simple models. First we will see how it appears in a $\Lambda$-CDM
model, that is, a Lemaître-Tolman-Bondi dust model with a cosmological
constant. We will then look at more general models including perfect-fluids.

\subsection{Overdensity in a $\Lambda$-CDM model\label{sub:Overdensity-in-a}}

In what follows we consider a $\Lambda$-LTB model which, besides
the bare LTB case, is exactly solvable, the simplest perfect-fluid
model with a cosmological context departing from LTB and which satisfies
the conditions for the existence of an asymptotically $r$-static
dividing shell. Indeed,%
{} as stated in \citep{LaskyLun06b}, choosing $P=0$ leads to the usual
LTB solutions. Setting $P=0$ in Eq.~(\ref{eq:MdotLTB}) implies%
\footnote{$M$ can be understood as the mass of the dust alone but interacting
with $\Lambda$, see Sec. \ref{sub:Remarks-on-Lambda}.%
} $\dot{M}=0$, and it is somewhat remarkable that this mass is still
conserved for each shell in spite of the presence of $\Lambda$. $\Lambda$
gives a homogeneous pressure, which in Eq.~(\ref{eq:EulerLTB}) gives
$\alpha^{\prime}=0$ so we can redefine $\alpha dT=dT^{*}$ into the
line element (\ref{eq:dsLTB}), and finally in Eq.~(\ref{eq:EdotLTB}),
assuming no shell crossing $r^{\prime}\ne0$.%
{} We are therefore left with Eq.~(\ref{eq:RadEvolLTB}) in the classic
LTB form, with%
{}\begin{align}
\dot{r}^{2}= & 2\frac{M}{r}+\frac{1}{3}\Lambda r^{2}+E.\label{eq:RadLevolLCDM}\end{align}
 Adding a cosmological constant modifies the mass definition but not
the dust equation of motion. However, %
{}we have an extra term that leads to a different dynamics. We can thus
write the Raychaudhuri-like equation corresponding to time derivation
of Eq.~(\ref{eq:RadLevolLCDM}):\begin{align}
\ddot{r}= & -\frac{M}{r^{2}}+\frac{\Lambda}{3}r,\label{eq:Ray-like}\end{align}
and this shows there exists a radius without acceleration for strictly
positive $\Lambda$, contrary to pure dust. However, the first integral
(\ref{eq:RadLevolLCDM}) suffices for analysis of what happens to
each shell (with fixed $R$).

\subsubsection{Kinematic analysis\label{sub:Kinematic-analysis}}

The Friedmann-like Eq. (\ref{eq:RadLevolLCDM}) can be used to get
the dynamics in a purely kinematical way. It can be expressed with
a polynomial\begin{align}
\dot{r}^{2}= & \frac{\Lambda}{3r}\left(r^{3}+\frac{3E}{\Lambda}r+\frac{6M}{\Lambda}\right)=\frac{\Lambda}{3r}P_{3,f}(r),\label{eq:Friedm}\end{align}
 which roots (given in Appendix \ref{sec:Appendix-A:-Roots}) should
obey the effective potential equation\begin{align}
E=V(r)\equiv & -\frac{2M}{r}-\frac{\Lambda}{3}r^{2}.\label{eq:effectivePot}\end{align}
 Since $\dot{r}^{2}\ge0$, we have the condition\begin{align}
E\ge & V(r).\end{align}
 The motion of a given shell over time thus follows $E=\mathrm{const}$
curves above the effective potential $V$. Roots, the points of changing
direction, translate as geometric intersections between those curves
and $V$. The effective potential admits one real negative root (0
energy/curvature) at\begin{align}
r= & -\sqrt[3]{\frac{6M}{\Lambda}},\end{align}
 and one double solution at its horizontal tangent ($V^{\prime}=0$)\begin{align}
r_{\mathrm{lim}}= & \sqrt[3]{\frac{3M}{\Lambda}},\end{align}
 for which the value of $E$ becomes\begin{align}
E_{\mathrm{lim}}= & -\left(3M\right)^{2/3}\Lambda^{1/3}.\label{eq:Elim}\end{align}
It can easily be shown that any shell standing at $r_{\mathrm{lim}}$
with $E_{\mathrm{lim}}$ will automatically be a limit shell\begin{align}
r_{\mathrm{lim}}= & -\frac{2M_{\mathrm{tot,lim}}}{E_{\mathrm{lim}}}=-2\frac{M+\frac{\Lambda}{6}r_{\mathrm{lim}}^{3}}{E_{\mathrm{lim}}}=-\frac{3M}{E_{\mathrm{lim}}},\end{align}
and calculating its $\mathrm{gTOV}$, using the definition of Eq.~(\ref{eq:TOVdef})
and recognizing %
\begin{figure}
\selectlanguage{british}%
\includegraphics[bb=146bp 232bp 821bp 683bp,width=1\columnwidth]{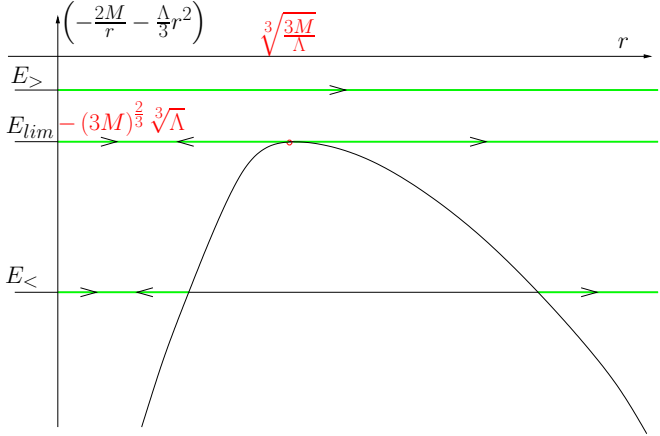}

\selectlanguage{english}%
\caption{\selectlanguage{british}%
\label{fig:kinematic-analysis}\foreignlanguage{english}{Kinematic
analysis for a given shell of constant $M$ and $E$. Depending on
$E$ relative to $E_{\mathrm{lim}}$, the fate of the shell is either
to remain bound ($E_{<}<E_{\mathrm{lim}}$) or to escape and cosmologically
expand ($E_{>}>E_{\mathrm{lim}}$). There exists a critical behavior
where the shell will forever expand, but within a finite, bound radius
($E=E_{\mathrm{lim}}$, $r\le r_{\mathrm{lim}}$). The maximum occurs
at $r_{\mathrm{lim}}=\sqrt[3]{3M/\Lambda}$.}\selectlanguage{english}
}

\end{figure}
Eq.~(\ref{eq:Ray-like}), \begin{align}
\mathrm{gTOV}= & \frac{M}{r^{2}}-\frac{\Lambda}{3}r=-\ddot{r},\label{eq:gTovLCDM}\end{align}
that such a shell will be $r$-static ($\mathrm{gTOV}_{\mathrm{lim}}=-\ddot{r}_{\mathrm{lim}}=0$).
The effective potential analysis is shown in Fig.~\ref{fig:kinematic-analysis}.

\begin{figure}
\includegraphics[width=1\columnwidth]{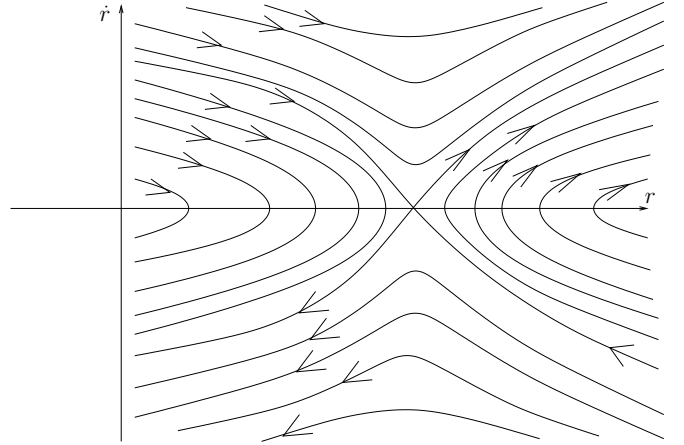}

\caption{\label{fig:phase-space}Phase space of a shell of fixed $M$ and $E$.
The scales are set by the value of $r_{lim}=\sqrt[3]{3M/\Lambda}$
while the actual kinematic of the shell is given by $E$.}

\end{figure}

We can thus reconstruct the phase space of that shell in the $(\dot{r},r)$
plane. Above the energy $E_{\mathrm{lim}}$, there is only one root
in the negative region; thus the flow is qualitatively defined by
its initial conditions. At $E_{\mathrm{lim}}$, the double positive
root gives a repulsive point, thus a saddle, while, below $E_{\mathrm{lim}}$,
the pair of roots give closed and open orbits as shown in Fig.~\ref{fig:phase-space}.

The Raychaudhuri-like equation can also be expressed with a polynomial\begin{align}
\ddot{r}= & \frac{\Lambda}{3r^{2}}\left(r^{3}-\frac{3M}{\Lambda}\right)=\frac{\Lambda}{3r^{2}}P_{3,R}(r),\end{align}
 admitting only one real root; the acceleration is always positive
for\begin{align}
r\ge & \sqrt[3]{\frac{3M}{\Lambda}},\end{align}
 thus at infinity (cosmological constant dominates, and $M$ is monotonous
in $r$). Therefore, at this root, there exists a limit radius beyond
which there is no recollapse:\begin{align}
r_{\mathrm{lim}}(R)= & \sqrt[3]{\frac{3M(R)}{\Lambda}}.\end{align}
 Note that this radius corresponds to the saddle point, which initial
energy radial profile is fixed with initial conditions for the mass
distribution $E_{\mathrm{lim}}(R)=-\left(3M(R)\right)^{2/3}\Lambda^{1/3}.$
Therefore the last intersection between the initial curvature profile,
set by combining velocity and mass profiles, and this saddle point
profile yields a global shell beyond which there is no recollapse,
recovering separation of expansion from collapse. Explicit exact solutions
for this $\Lambda$LTB evolution model are shown in Appendix \ref{sec:Appendix-B:-Exact}.
It is nevertheless crucial to realize that the selection of the limit
shell from initial curvature does not entail necessarily that it should
start as $r$-static. Indeed the opposite should be true in general,
as can be seen in Eqs.~(\ref{eq:RadLevolLCDM}) using $E_{\mathrm{lim}}$,
$R_{\mathrm{lim}}$ in (\ref{eq:effectivePot}), and Fig.~\ref{fig:kinematic-analysis}:
for any choice of the initial $R_{\mathrm{lim}}<r_{\mathrm{lim}}$,
the radial velocity\begin{align}
\dot{R}_{\mathrm{lim}}^{2}= & E_{\mathrm{lim}}-V(R_{\mathrm{lim}})>0,\end{align}
so it appears that the $r$-static behavior of the shell should only
emerge asymptotically as it approaches zero velocity for infinite
time. The selected limit shell therefore agrees with the conditions~(\ref{eq:LimitDef},\ref{eq:Lie2rStar})
only at infinity in time, and is traced back to initial conditions
owing to the $\Lambda$+dust conservation of $M$ and $E$ in time.
More general fluids should not always allow for this conservation
on the limit shell; however, once a shell verifies Eqs.(\ref{eq:LimitDef},\ref{eq:Lie2rStar}),
its staticity guarantees that it should verify it at time infinity.
It is remarkable that the existence of the limit shell only matters
at time infinity, suggesting that a weaker definition than~(\ref{eq:LimitDef})
and (\ref{eq:Lie2rStar}), should be a sufficient condition.

\subsubsection{Time dependent $\mathrm{TOV}$}

The shape of Eq.~(\ref{eq:gTovLCDM}) shows that, at the root of
the Raychaudhuri-like polynomial, $\mathrm{gTOV}=0$ and that it is
positive inside and negative outside. The trapped region is thus characterized
by $\mathrm{gTOV}\ge0$. We can also compute, using $M=4\pi\left\langle \rho\right\rangle r^{3}/3$,\begin{align}
\mathrm{gTOV}^{\prime}= & \left[4\pi\left(\rho-\frac{2}{3}\left\langle \rho\right\rangle \right)-\frac{\Lambda}{3}\right]r^{\prime}\end{align}
 so $\mathrm{TOV}$ is a decreasing function of $r$ {[}for $r^{\prime}>0$,
a fair assumption as seen when $r(t=0)=R${]}, except in regions where
$\rho>\frac{2}{3}\left(\left\langle \rho\right\rangle +\rho_{\Lambda}\right)$,
that is, in density peaks. It is also a time dependent function through
the evolution of $r$:\begin{align}
\dot{\mathrm{gTOV}}= & \mp\left(\frac{2M}{r^{3}}+\frac{\Lambda}{3}\right)\sqrt{E+\frac{2M}{r}+\frac{\Lambda}{3}r^{2}},\end{align}
and thus for a given shell, it increases with time for ingoing dust
shells and decreases for outgoing ones. The main point is that with
dust, turnaround shells have $r$-static $\mathrm{gTOV}$, and that
balanced shells (between their mass pull and that of $\Lambda$) verify
the $\mathrm{TOV}$ equation and are thus static.

\subsubsection{Expansion and shear}

\begin{figure}
\includegraphics[width=0.8\columnwidth]{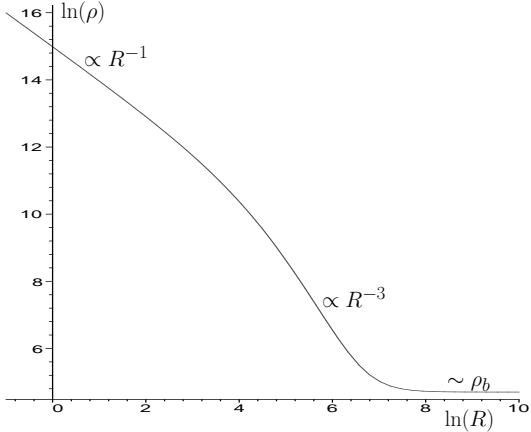}

\caption{\label{cap:NFW+bDen}NFW with background density profile}

\end{figure}
From the definition (\ref{eq:aMetric1}) of the shear, we see that
in the GLTB model under consideration \begin{equation}
a=-\frac{1}{3}\left(\frac{\dot{r}^{\prime}}{r^{\prime}}-\frac{\dot{r}}{r}\right)\;,\end{equation}
 where we now denote by a prime the derivative with respect to the
GLTB radial coordinate $R$ (i.e., $\partial r=\partial R/r^{\prime}$).
Using Eqs.~(\ref{eq:RadLevolLCDM}) and (\ref{eq:Ray-like}) we,
then, derive \begin{multline}
\hspace{-0.5cm}a=\mp\frac{1}{6\,\sqrt{{\scriptstyle E+2\frac{M}{r}+\frac{\Lambda}{3}r^{2}}}}\left[\left(\frac{E^{\prime}}{r^{\prime}}-\frac{2E}{r}\right)+\frac{2}{r}\,\left(\frac{M^{\prime}}{r^{\prime}}-\frac{3M}{r}\right)\right].\end{multline}
 It is then possible to verify that this quantity does not vanish
in general when $r\to r_{\star}$. It does vanish if the expansion
$\Theta$ also vanishes at the locus where $\beta/\alpha=0$, i.e.,
at $r=r_{\star}$, as we have commented in Sec.~\ref{sub:Remarks-on-null}.
%
{}

\subsubsection{Examples of initial density\label{sub:Examples-of-initial}}

It is obvious then that initial conditions are crucial to determine
the existence of a separating shell in the $\Lambda$LTB model since
they set the profile of $E$ and that of $E_{\mathrm{lim}}$. A single
crossing of the two curves ensures locally the existence of such a
shell, while its global effect remains if the initial conditions do
not foster shell crossing. This is the case if there is only one crossing
from bound to unbound $E$ of $E_{\mathrm{lim}}$. More complicated
cases will be examined in a future work%
{}. We now proceed with examples of initial density profiles and then
deduct the conditions on the corresponding curvature profile for a
limit shell to exist.

\paragraph{NFW with background:}

The choice of a so-called Navarro, Frenk and White (NFW) density profile
\citep{NFW} is motivated by their prevalence in large cosmological
dark matter haloes (\citet{LeDPhD}, and references therein). If we
initialize the halo with such a density profile, with concentration
$1/R_{0}$ and inflection density $\rho_{0}/4$, placed on a constant
background $\rho_{b}$, we can compute the corresponding mass profile.
The density profile, as illustrated in %
\begin{figure}
\includegraphics[width=0.8\columnwidth]{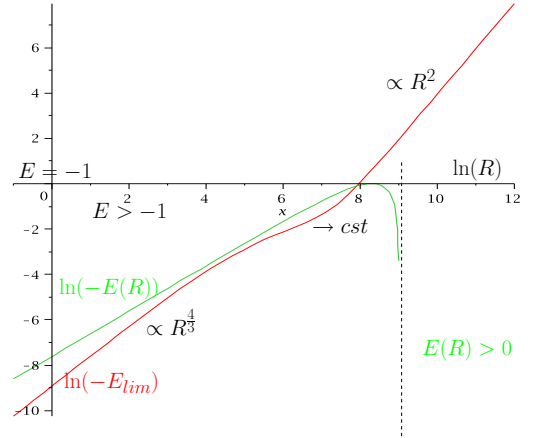}

\caption{\label{cap:NFW+bEmax}\label{cap:NFW+bE+Emax}NFW with background
$E_{\mathrm{lim}}$ and an example of $E$ profile given by Eq.~(\ref{eq:EprofExpl}),
for $E_{\mathrm{min}}=-1+e^{-10}$ and $r_{1}=e^{9}$.}

\end{figure}
Fig.~\ref{cap:NFW+bDen}, is given by \citep{NFW}\begin{align}
\rho= & \frac{\rho_{0}}{\frac{R}{R_{0}}\left(1+\frac{R}{R_{0}}\right)^{2}}+\rho_{b}.\end{align}
The corresponding mass then reads\begin{multline}
M=4\pi\left\{ r_{0}^{3}\rho_{0}\left[\ln\left(1+\frac{R}{r_{0}}\right)-\frac{R}{R+r_{0}}\right]+\rho_{b}\frac{R^{3}}{3}\right\} .\end{multline}
 Now armed with the expression for the maximum energy function, the
double root solution above, we can obtain from Eq.~(\ref{eq:Elim})
the bound upper limit for the initial energy/curvature profile that
separates between ever-expanding and bound shells\begin{multline}
E_{\mathrm{lim}}=-\left(12\pi\right)^{2/3}\Lambda^{1/3}\left\{ r_{0}^{3}\rho_{0}\left[\ln\left(1+\frac{R}{r_{0}}\right)-\frac{R}{R+r_{0}}\right]\right.\\
\left.+\rho_{b}\frac{R^{3}}{3}\right\} ^{2/3}.\end{multline}
 Figure~\ref{cap:NFW+bEmax} shows that profile corresponding to
the NFW with background mass. We then propose an example for the $E(R)$
profile, motivated by its cosmological Friedmann asymptotic curvature
and its simple radial evolution from bound to unbound, as\begin{align}
E(R)= & -4E_{\mathrm{min}}\left(\frac{R}{r_{1}}\right)\left(1-\frac{R}{r_{1}}\right),\label{eq:EprofExpl}\end{align}
 where $r_{1}>0$ and $-1<E_{\mathrm{min}}<0$, chosen so that $E$
crosses $E_{\mathrm{lim}}$ near its constant density region. With
the asymptotic constant density and Friedmann negative curvature ($E\simeq\frac{4}{r_{1}^{2}}R^{2}=-k_{\infty}R^{2}$),
these initial conditions model well a collapsing structure in an open
background of curvature radius $\frac{r_{1}}{2}$. The resulting curves
are shown in Fig.~\ref{cap:NFW+bE+Emax}. We have here an example
where shells with $E<E_{\mathrm{lim}}$ are trapped inside the limit
shell defined by the intersection of the two profiles. Moreover, that
limit shell in the case of dust with $\Lambda$ has been shown to
be static. Thus, with this set of physically motivated initial conditions,
the limit shell defined in this way delimits a constant region of
collapsing mass, separated from expanding shells.

\paragraph{Cosmological background with power law overdensity:}

\begin{figure}
\includegraphics[width=0.8\columnwidth]{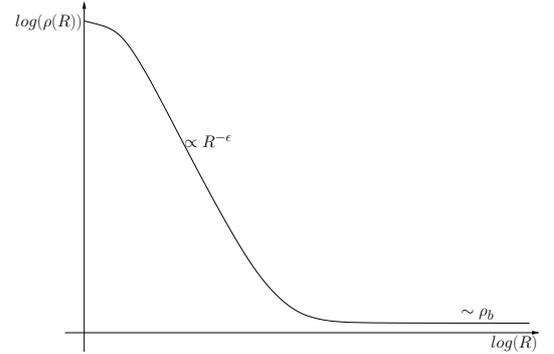}

\caption{\label{cap:Eps+bDen}Power law density profile without cusp and with
background}

\end{figure}

The most natural cosmological initial condition is a power law overdensity,
with or without cusp, upon a uniform background with an initial Hubble
flow (\citet{LeDPhD}). The uniform background and initial Hubble
flow ensures the asymptotic solution starts FLRW. In this second example
of initial conditions, we explored both density profiles but illustrate
only the cuspless case as it is more observationally sounded (\citet{LeDPhD},
and references therein). The density profiles, as illustrated for
the second case in Fig.~\ref{cap:Eps+bDen}, are given by ($\epsilon>0$,
and in the %
first case $\epsilon\le3$ for a finite central mass)\begin{align}
\rho= & \rho_{0}\left(\frac{R}{R_{0}}\right)^{-\epsilon}+\rho_{b},\\
\rho= & \rho_{0}\left(1+\frac{R}{R_{0}}\right)^{-\epsilon}+\rho_{b}.\end{align}
Observations of the cosmic microwave background would imply the choice
of initial time at recombination and amplitudes of the order of $\rho_{0}\sim10^{-5}\rho_{b}$
(see \citet{LeDPhD}, and references therein). The corresponding mass
then reads, for the cuspy profile,\begin{multline}
M_{cusp}=4\pi r_{0}^{3}\rho_{0}\left\{ \begin{array}{ll}
\left[\ln\left(\frac{R}{r_{0}}\right)\right], & \epsilon=3\\
\left[\frac{\left(\frac{R}{r_{0}}\right)^{3-\epsilon}}{3-\epsilon}\right], & 0<\epsilon<3\end{array}\right\} +\frac{4\pi}{3}\rho_{b}R^{3},\end{multline}
 and for the profile with constant density in the center%
\begin{figure}
\includegraphics[width=0.8\columnwidth]{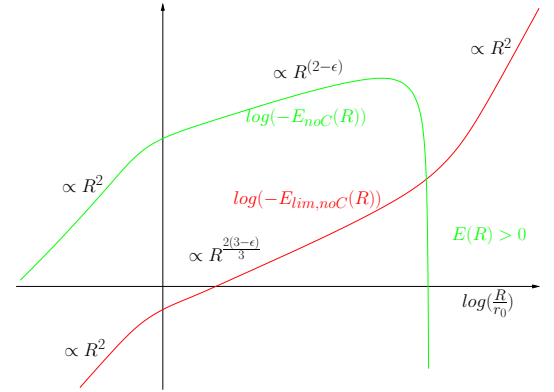}

\caption{\label{cap:Eps+bE+Emax}Power law density without cusp + background
in$\log(-E_{\mathrm{lim}})-\log(R)$ and $\log(-E)-\log(R)$ scales}

\end{figure}
\begin{widetext}\begin{multline}
M_{\mathrm{no\, Cusp}}=4\pi r_{0}^{3}\rho_{0}\times\left\{ \begin{array}{ll}
\left[\frac{1}{2}\left(\frac{R}{r_{0}}\right)\left(\frac{R}{r_{0}}-2\right)+\ln\left(1+\frac{R}{r_{0}}\right)\right], & \epsilon=1\\
\left[\left(\frac{R}{r_{0}}\right)\frac{2+\frac{R}{r_{0}}}{1+\frac{R}{r_{0}}}-2\ln\left(1+\frac{R}{r_{0}}\right)\right], & \epsilon=2\\
\left[\frac{\frac{R}{r_{0}}}{\left(1+\frac{R}{r_{0}}\right)^{2}}+\ln\left(1+\frac{R}{r_{0}}\right)\right], & \epsilon=3\\
\left[\frac{\left(1+\frac{R}{r_{0}}\right)^{3-\epsilon}-1}{3-\epsilon}-2\frac{\left(1+\frac{R}{r_{0}}\right)^{2-\epsilon}-1}{2-\epsilon}+\frac{\left(1+\frac{R}{r_{0}}\right)^{1-\epsilon}-1}{1-\epsilon}\right], & \epsilon>0\end{array}\right\} +\frac{4\pi}{3}\rho_{b}R^{3}.\end{multline}
\end{widetext} The resulting boundary profile for $E$ again follows
Eq.~(\ref{eq:Elim}), using the obtained mass profiles. Taking an
initial Hubble flow, $\dot{R}=H_{i}R$, the $E(R)$ profile is then
defined by Eq.~(\ref{eq:RadLevolLCDM}) to be\begin{align}
E(R)= & \left(H_{i}^{2}-\frac{\Lambda}{3}\right)R^{2}-\frac{2M}{R}.\end{align}
The resulting comparison between $E$ and $E_{lim}$ for the noncuspy
case is shown in Fig.~\ref{cap:Eps+bE+Emax}. Once again, the intersection
defines a static limit shell for which $r_{\mathrm{lim}}=-\frac{2M_{\mathrm{tot,\, lim}}}{E_{\mathrm{lim}}}$
and $\mathrm{gTOV}=0$, all shells inside it are in the kinematically
bound region of Fig.~\ref{fig:kinematic-analysis}, while those outside
are in the free region. Initial conditions ensure they will expand
in a quasi-FLRW manner. 

These examples illustrate that cosmologically motivated initial conditions
lead to a clear separation between expanding and collapsing regions.
Therefore for these systems, expansion ignores the effects of collapse,
and conversely the details of the collapsing region can ignore the
presence of a background expanding universe.


\subsection{perfect-fluid core in a $\Lambda$-CDM model}

Before examining the possibility of existence for a limit shell inside
a perfect-fluid in a sequel paper%
{}, where we shall present an ansatz for a perfect-fluid inhomogeneous
core in a Friedmann environment, let us turn to the configuration
where a perfect-fluid ball is surrounded by (a) vacuum with a cosmological
constant, and (b) dust and $\Lambda$.

\subsubsection{Pure $\Lambda$ exterior}

In the same way as \citep{LaskyLun06b} did for a perfect-fluid surrounded
by a $\Lambda=0$ vacuum, we can examine the interface between the
perfect-fluid and the $\Lambda$ vacuum. In the latter region, both
\emph{the pressure radial derivative $P^{\prime}=0$ and the sum $\rho_{\Lambda}+P_{\Lambda}=0$
for all time and place by definition of $\Lambda$}. In the same way
as \citep{LaskyLun06b} showed for such a configuration with $\Lambda=0$
vacuum, such a simple interface implies, through Eqs.~(\ref{eq:EulerLTB})
and (\ref{eq:LieE}), that the energy and lapse functions, $E$ and
$\alpha$, are undefined there. These equations show that only if
the fluid's pressure radial derivative $P^{\prime}$ vanishes faster
than $\rho+P$ can $E$ and $\alpha$ remain defined. This condition
sets an unusual boundary constraint to the perfect-fluid's EoS (simple
linear EoS do not agree with it), but it is more fruitful to point
out that such behavior mimics that of a vanishingly thin layer of
$\Lambda$-dust. Thus, the transition between the two regimes gives
rise to an inescapable $\Lambda$-dust atmosphere, however vanishingly
thin, as was found in the pure vacuum case \citep{LaskyLun06b}. We
have two free boundaries, $r_{\partial_{1}}(t)$ where the pressure
vanishes and $r_{\partial_{2}}(t)>r_{\partial_{1}}(t)$ where the
density vanishes, at which the EoS is %
\begin{figure}
\includegraphics[width=0.8\columnwidth]{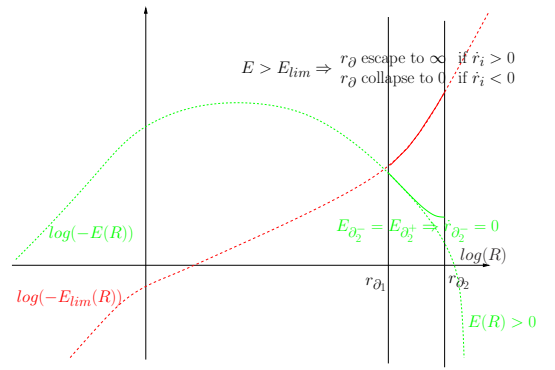}

\caption{\label{fig:-case-for<}$r_{\mathrm{lim}}<r_{\partial_{1}}<r_{\partial_{2}}$
case for a dust layer with $\Lambda$. Full space $\Lambda$-CDM diagram
for $\log(-E_{\mathrm{lim}})-\log(R)$ and $\log(-E)-\log(R)$ in
dashed line. This region is characterized by $E>E_{\mathrm{lim}}$,
so the dynamical analysis of Fig.~\ref{fig:kinematic-analysis} yields
continuation of initial velocities directions.}

\end{figure}
defined as\begin{align}
0= & \begin{cases}
f(\rho,P) & \textrm{for }r\in\left[0;r_{\partial_{1}}\right]\\
P & \textrm{for }r\in\left[r_{\partial_{1}};r_{\partial_{2}}\right].\end{cases}\end{align}
Evolution of $r_{\partial_{1}}(t)$ and $r_{\partial_{2}}(t)$ follows
from setting, respectively, $P=0$, then $P=\rho=0$ in Eqs.~(\ref{eq:Mdot}),
(\ref{eq:Edot}), and (\ref{eq:EulerLTB}), to evolve those radii
from initial conditions. The continuity of the curvature through both
boundaries imposes again\begin{align}
\left[\lim_{r\rightarrow r_{\partial_{i}}^{+}}-\lim_{r\rightarrow r_{\partial_{i}}^{-}}\right]\left\{ E\left(t,r\right)\right\} = & 0,\label{eq:curvContin}\end{align}
which can be used to transmit the value of the mass parameter from
the outer Schwarzschild-de~Sitter spacetime down to the perfect-fluid
boundary curvature.

\subsubsection{Limit shell}

\begin{figure}
\includegraphics[width=0.8\columnwidth]{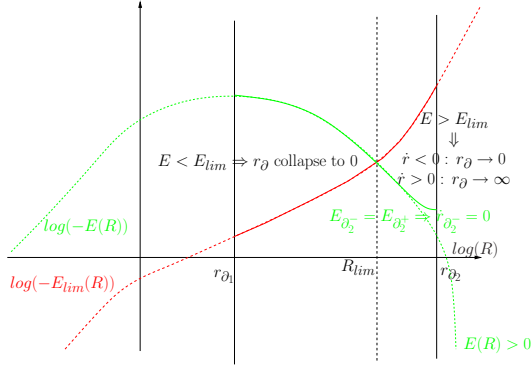}

\caption{\label{fig:-case-forI}$r_{\partial_{1}}<r_{\mathrm{lim}}<r_{\partial_{2}}$
case for a dust layer with $\Lambda$, $\Lambda$-CDM for $\log(-E_{\mathrm{lim}})-\log(R)$
and $\log(-E)-\log(R)$ in dashed line. The region with $E<E_{\mathrm{lim}}$
is trapped by its set of effective potentials and will recollapse
that with $E>E_{\mathrm{lim}}$, so the dynamical analysis of Fig.~\ref{fig:kinematic-analysis}
yields continuation of initial velocities. The separating shell remains
in between those regions.}

\end{figure}
At this stage, the possibility opens for a limit shell in the $\Lambda$-CDM
atmosphere of the core, provided that such a shell verifies in conjunction
Eqs.~(\ref{eq:LimitShell}), or equivalently (\ref{eq:LimitDef}),
and (\ref{eq:Lie2rStar}), which is only possible in a positively
curved region. Given the surrounding Schwarzschild-de~Sitter environment,
the positive curvature requirement is at least locally filled near
the outer boundary. There the analysis of Sec. (\ref{sub:Overdensity-in-a})
applies fully to yield, given initial conditions, the location of
the previously discussed static virtual shell. Recall that in the
Schwarzschild-de~Sitter region, $E=-\frac{2M_{\partial_{2}}}{r}-\frac{\Lambda}{3}r^{2}$
while $E_{\mathrm{lim}}=-\left(3M_{\partial_{2}}\right)^{2/3}\Lambda^{1/3}=cst$;
however, the analysis only applies in the presence of dust, thus between
$r_{\partial_{1}}$ and $r_{\partial_{2}}$. Owing to the preservation
of continuity in $M$ and $E$ at $r_{\partial_{1}}$, whichever behavior
the perfect-fluid may have, it will be confined by that of the previously
explored $\Lambda$-CDM at its boundary.

\begin{figure}
\includegraphics[width=0.8\columnwidth]{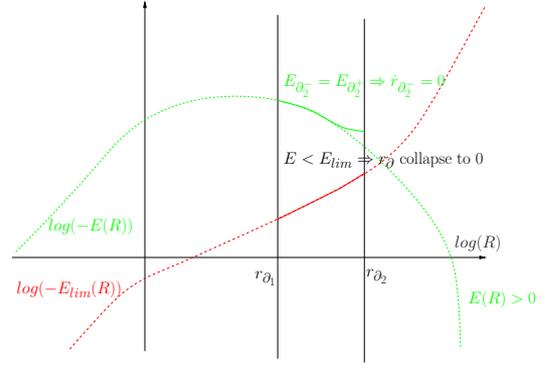}

\caption{\label{fig:-case-for>}$r_{\partial_{1}}<r_{\partial_{2}}<r_{\mathrm{lim}}$
case for a dust layer with $\Lambda$, $\Lambda$-CDM for $\log(-E_{\mathrm{lim}})-\log(R)$
and $\log(-E)-\log(R)$ in dashed line. This region is characterized
by $E<E_{\mathrm{lim}}$, so the dynamical analysis of Fig.~\ref{fig:kinematic-analysis}
yields eventual recollapse.}

\end{figure}

Let us exhibit examples of such configurations: we can start from
a similar example as presented in Sec. (\ref{sub:Examples-of-initial}).
Nevertheless, to preserve curvature continuity (\ref{eq:curvContin}),
the initial velocity at $r_{\partial_{2}}$ should go to 0, and therefore
the previous $E$ profile should be modified accordingly. Then we
are faced with three possibilities due to the location of the dust
layer boundaries compared with the limit shell in the full space dust
model: $r_{\mathrm{lim}}<r_{\partial_{1}}<r_{\partial_{2}}$, $r_{\partial_{1}}<r_{\mathrm{lim}}<r_{\partial_{2}}$,
or $r_{\partial_{1}}<r_{\partial_{2}}<r_{\mathrm{lim}}$. Those cases
are illustrated, respectively, in Figs. \ref{fig:-case-for<}, \ref{fig:-case-forI}
and \ref{fig:-case-for>}. In the first case, the dust layer locates
above the maximum of their effective potential (\ref{eq:effectivePot})
so their initial velocities gives the direction of their unhindered
asymptotic behavior; i.e. an initially expanding dust layer should
expand forever. If a separating shell exists, it should lie within
the perfect-fluid region. The second case shows the existence of a
separating shell, the perfect-fluid being bound by the eventual recollapse
of the $r_{\partial_{1}}$ shell, while some of the dust shell will
expand through the vacuum region and eventually squeeze it to infinity.
In the third case all the dust shells locate below the maximum of
their effective potential (\ref{eq:effectivePot}) so the whole mass
will eventually recollapse, as if the separating shell was virtually
located in the vacuum region.

Now sending the $r_{\partial_{2}}$ boundary to infinity, we can expand
the dust layer accordingly and so long as Sec. (\ref{sub:Overdensity-in-a})'s
analysis yields a limit shell within the dust region, the perfect-fluid
shall be contained by the collapsing inner boundary (i.e. the third
case disappear and we are left with cases $r_{\mathrm{lim}}<r_{\partial_{1}}$
and $r_{\partial_{1}}<r_{\mathrm{lim}}$ as treated in Figs.~\ref{fig:-case-for<}
and \ref{fig:-case-forI}).

In this section we have found that the presence of a cosmological
constant does not modify the need for a dust layer around a perfect-fluid
core surrounded by vacuum. We have also given examples of limit shell
separation behaviors for appropriately set initial conditions in the
dust layer with $\Lambda$. We have even hinted at that possibility
inside the perfect-fluid from the dust behavior, although such study
should be left for a sequel paper.%
{}

%
{}

\section{Summary and discussion}

In the present work we have considered spherically symmetric, inhomogeneous
universes in order to ascertain under which conditions a dividing
shell separating expanding and collapsing regions exists. This endeavor
is important in relation with the present understanding of structure
formation as the outcome of gravitational collapse of overdense patches
within an overall expanding universe%
{}. 

We have addressed this problematic by resorting to an ADM 3+1 splitting,
utilizing the so-called generalized Painlevé-Gullstrand coordinates
as developed in Refs.~\citep{Adler et al 05,LaskyLun06b}. This enables
us to follow a nonperturbative approach and to avoid having to consider
the matching of the two regions with the contrasting behaviors \citep{Matravers:2000cu}.
We have found local conditions characterizing the existence of a dividing
shell. %
{} We have related these conditions to a gauge invariant definition
of the properties of the dividing shell. These require the vanishing
of a linear combination of the expansion scalar and of the shear on
the shell, as well as that of its flow derivative. In GPG coordinates,
it summarizes as a vanishing of both first- and second-order flow
derivatives of the areal radius. 

%
{}In order to illustrate our findings we have considered some simple
examples of cosmological interest that provide realizations of our
results. We have considered a $\Lambda$-CDM model whereby we consider
an LTB universe with dust and a cosmological constant. Notice that
the simultaneous consideration of the latter two components yields
a perfect-fluid model for the combined matter content. Moreover it
can be seen as a simplified model of a dust universe within a cosmological
setting coarsely provided by $\Lambda$, which would then mimic the
energy content of the background cosmological model with a rate of
expansion much smaller than that of the pure dust collapse.

We have chosen initial conditions motivated by cosmological considerations
and have discussed the existence of a dividing shell for those cases.
We have also generalized a result of Ref.~\citet{LaskyLun06b} for
the case where a cosmological constant is present, which states that
a perfect-fluid core embedded in a universe filled with a cosmological
constant necessarily exhibits a dust transition between the perfect-fluid
inner region and the outer vacuum region. This permits one to envisage
this case as a generalization of the former $\Lambda$-CDM examples.

Finally we should mention that a thorough discussion of global conditions
represents a much harder problem, and remains an open problem since
this involves the full characterization of a partial differential
equations problem with boundary conditions in an open domain. %
{}

\begin{acknowledgments}
The authors wish to thank José Fernando Pascual-Sanchez for bringing
to their attention the work of the authors of Ref.~\citep{LaskyLun06b}
and for helpful discussions. The work of M.Le~D. is supported by
CSIC (Spain) under Contract No. JAEDoc072, with partial support from
CICYT Project No. FPA2006-05807, at the IFT, Universidad Autonoma
de Madrid, Spain, and was also supported by FCT (Portugal) under Grant
No. SFRH/BPD/16630/2004, at the CFTC, Lisbon University, Portugal.
F.C.M. is supported by CMAT, Univ. Minho, FCT Project No. PTDC/MAT/108921/2008
and Grant No. SFRH/BSAB/967/2010. J.P.M. also wishes to thank the
provider of Grants No. PTDC/FIS/102742/2008 and No. CERN/FP/109381/2009.
\end{acknowledgments}
\appendix

\section{Roots of $P_{3,f}(r)$\label{sec:Appendix-A:-Roots}}

\subsection{Roots for the polynomial}

The roots ($r_{0}$) of Eq.~(\ref{eq:Friedm}) proceed from the polynomial
$P_{3,f}$. We change variables such that $r=u+v$ and use the extra
degree of freedom to choose to rewrite $P_{3,f}=0$ such that\begin{align}
uv= & -\frac{E}{\Lambda},\label{eq:uvI}\\
\left(u^{3}+\frac{3M}{\Lambda}\right)^{2}= & \left(\frac{E}{\Lambda}\right)^{3}+\left(\frac{3M}{\Lambda}\right)^{2}.\end{align}
 Solutions for the latter second degree polynomial come naturally
as\begin{align}
u^{3}= & \frac{-3M\pm\sqrt{\frac{E^{3}}{\Lambda}+\left(3M\right)^{2}}}{\Lambda}\\
\Rightarrow u= & \sqrt[3]{\frac{-3M\pm\sqrt{\frac{E^{3}}{\Lambda}+\left(3M\right)^{2}}}{\Lambda}}e^{i(2\pi k/3)}.\end{align}
 We are left with six solutions for $u$ and $v$, which are symmetrical
and related by Eq.~(\ref{eq:uvI}) so $uv$ being real, choosing
$u^{3}$ as the positive square root solution, the corresponding $v^{3}$
becomes the negative one while $u$ and $v$ are complex conjugate,
so \begin{align}
uv= & \sqrt[3]{\frac{\left(3M\right)^{2}-\frac{E^{3}}{\Lambda}-\left(3M\right)^{2}}{\Lambda^{2}}}=-\frac{E}{\Lambda},\end{align}
 and therefore the roots are\begin{align}
r_{k=0,\pm1}= & \left(\sqrt[3]{-3M+\sqrt{\frac{E^{3}}{\Lambda}+\left(3M\right)^{2}}}e^{i(2\pi k/3)}\right.\nonumber \\
 & \left.+\sqrt[3]{-3M-\sqrt{\frac{E^{3}}{\Lambda}+\left(3M\right)^{2}}}e^{-i(2\pi k/3)}\right)/\Lambda^{1/3}.\label{eq:RkSolnsI-1}\end{align}

\subsection{Real root(s)}

For the positive discriminant, $\Delta=\frac{E^{3}}{\Lambda}+\left(3M\right)^{2}$,
there is only one real root for $k=0$. A negative or null discriminant
yields again the real $k=0$ root and two other real roots for $k=\pm1$,
since then $v=\overline{u}$. We are then left with the single real
root, noting \begin{align}
a_{0}= & \sqrt[3]{-3M+\sqrt{\frac{E^{3}}{\Lambda}+\left(3M\right)^{2}}},\\
a_{0}^{*}= & \sqrt[3]{-3M-\sqrt{\frac{E^{3}}{\Lambda}+\left(3M\right)^{2}}},\\
r_{0}= & \frac{a_{0}+a_{0}^{*}}{\Lambda^{1/3}},\end{align}
 and, when $\frac{E^{3}}{\Lambda}+\left(3M\right)^{2}\le0$, the two
other real roots\begin{align}
a_{\pm}= & \sqrt[3]{3M+i\sqrt{\frac{\left(-E\right)^{3}}{\Lambda}-\left(3M\right)^{2}}}(1\mp i\sqrt{3}),\\
\bar{a}_{\pm}= & \sqrt[3]{3M-i\sqrt{\frac{\left(-E\right)^{3}}{\Lambda}-\left(3M\right)^{2}}}(1\pm i\sqrt{3}),\\
r_{\pm}= & \frac{a_{\pm}+\bar{a}_{\pm}}{2\Lambda^{1/3}},\end{align}

\subsection{Signs of the real roots:}

So as to order the roots, it is necessary to look at their sign. This
is important as $r$ should be positive, $r<0$ being unphysical.
Recall that $M,\Lambda>0$, and $E>-1$. When $\Delta>0$, i.e. when
$E>-\left(3M\right)^{2/3}\Lambda^{1/3}=E_{\mathrm{lim}}$, we have
only one real root and $r_{0}>0\Rightarrow a_{0}>-a_{0}^{*}$. We
always have $-a_{0}^{*}=\sqrt[3]{3M+\sqrt{\frac{E^{3}}{\Lambda}+\left(3M\right)^{2}}}>0$.
Supposing $a_{0}>0$ (and thus $a_{0}^{3}>0$) then $-a_{0}^{*3}a_{0}^{3}=\frac{E^{3}}{\Lambda}>0\Leftrightarrow E>0$.
Therefore, with the hypothesis $E>0$, the condition $r_{0}>0$ implies
$a_{0}>-a_{0}^{*}\Leftrightarrow a_{0}^{3}>-a_{0}^{*3}\Leftrightarrow-3M>3M$.
Hence for $E>0$ we have $r_{0}<0$. The same as for $0\ge E>-\left(3M\right)^{2/3}\Lambda^{1/3}$,
requesting $r_{0}>0$ implies $a_{0}>-a_{0}^{*}$ while $-a_{0}^{*}>0\ge a_{0}$.
Therefore, $0\ge E>-\left(3M\right)^{2/3}\Lambda^{1/3}$ always entails
$r_{0}<0$, and we conclude that $r_{0}$ is always negative when
$E>-\left(3M\right)^{2/3}\Lambda^{1/3}$. The case when $\left(3M\right)^{2}\Lambda<1$
is more interesting as we have three real roots for $-1<E\le-\left(3M\right)^{2/3}\Lambda^{1/3}$.
Let us use the solutions of Eq.~(\ref{eq:RkSolnsI-1}) in the form\begin{align}
r_{k}= & \frac{u_{k}+\bar{u}_{k}}{\Lambda^{\frac{1}{3}}}=\frac{2\mathrm{Re}(u_{k})}{\Lambda^{\frac{1}{3}}}.\end{align}
 We know that\begin{align}
u_{k}^{3}= & -3M+i\sqrt{\frac{\left(-E\right)^{3}}{\Lambda}-\left(3M\right)^{2}}\end{align}
 so $\mathrm{Im}(u_{k}^{3})\ge0$ and $\mathrm{Re}(u_{k}^{3})<0$.
We can then rewrite $u_{k}^{3}=\rho e^{i\varphi_{k,3}}$ with $\rho^{2}=\frac{\left(-E\right)^{3}}{\Lambda}$,
and $\varphi_{k,3}\in\left[\begin{array}{ccc}
\frac{\pi}{2}+2k\pi & ; & \pi+2k\pi\end{array}\right]_{k\in\mathbb{Z}}$. The values of $u_{k}$ are deduced as $u_{k}=\rho^{\frac{1}{3}}e^{i\varphi_{k}}$
with $\varphi_{k}=\frac{\varphi_{k,3}}{3}$: $\varphi_{k}\in\left[\begin{array}{ccc}
\frac{\pi}{6}+\frac{2k\pi}{3} & ; & \frac{\pi}{3}+\frac{2k\pi}{3}\end{array}\right]_{k\in\mathbb{Z}}$. Each $u_{k}$ admits the same modulus, so the phases, each separated
by $2\pi/3$, give us the ranges and the order in which each root
lies. The results are the following:\begin{align}
\varphi_{0}\in & \left[\begin{array}{ccc}
\frac{\pi}{6} & ; & \frac{\pi}{3}\end{array}\right]\subset\left[\begin{array}{ccc}
0 & ; & \frac{\pi}{2}\end{array}\right] & \Rightarrow r_{0}> & 0,\\
\varphi_{+}\in & \left[\begin{array}{ccc}
\pi-\frac{\pi}{6} & ; & \pi\end{array}\right]\subset\left[\begin{array}{ccc}
\frac{\pi}{2} & ; & \pi\end{array}\right] & \Rightarrow r_{+}< & 0,\\
\varphi_{-}\in & \left[\begin{array}{ccc}
-\frac{\pi}{2} & ; & -\frac{\pi}{3}\end{array}\right]\subset\left[\begin{array}{ccc}
-\frac{\pi}{2} & ; & 0\end{array}\right] & \Rightarrow r_{-}\ge & 0,\end{align}
and the order of the cosine (since $r_{k}$ involves the real part
of $u_{k}$) yields $-r_{+}\ge r_{0}\ge r_{-}\ge0.$ This is agreeing
with the analysis of Sec. \ref{sub:Kinematic-analysis} understanding
that the negative root shifts from $r_{0}$ to $r_{+}$ through the
$\Delta=0$ point, and that below the horizontal tangent, $r_{0}$
is the exterior turning point while $r_{-}$ gives the interior envelope
of the effective potential.

The above solutions give us then the explicit equations for the intersection
of the effective potential with the current curvature involved in
Eq. \ref{eq:RadLevolLCDM}.

\section{Exact solutions for an inhomogeneous $\Lambda$CDM\label{sec:Appendix-B:-Exact}}

The equation of motion admits analytical solutions in terms of hyperelliptic
integrals%
{} (see also Lemaître \citep{Lemaitre33}). %
{}From Eq.~(\ref{eq:RadLevolLCDM})\begin{align}
t-t_{B}= & \int_{R}^{r}\sqrt{\frac{r}{Er+2M+\frac{\Lambda}{3}r^{3}}}dr;\end{align}
 however, in conformal time ($dt=rd\eta$)\begin{align}
r^{\prime2}= & Er^{2}+2Mr+\frac{\Lambda}{3}r^{4},\\
\Rightarrow\eta-\eta_{B}= & \int_{R}^{r}\frac{1}{\sqrt{Er^{2}+2Mr+\frac{\Lambda}{3}r^{4}}}dr\nonumber \\
= & \int_{R}^{r}\frac{1}{\sqrt{P_{4}(r)}}dr\end{align}
 Given that the incomplete elliptic integral of the first kind is
defined by\begin{align}
\! F(x,k)= & \int_{0}^{x}\frac{dt}{\sqrt{\left(1-t^{2}\right)\left(1-k^{2}t^{2}\right)}}=\int_{0}^{x}\frac{dt}{\sqrt{{\scriptstyle P_{F}\left(t\right)}}},\end{align}
 it is possible by a rational change of variable, $z=\frac{ax+b}{cx+d}$
to go from $P_{F}$ to $P_{4}$:\begin{align}
P_{F}\left(z(x)\right)= & \left((c-a)x+(d-b)\right)\left((c+a)x+(d+b)\right)\nonumber \\
 & \times\left((c-ka)x+(d-kb)\right)\nonumber \\
 & \times\left((c+ka)x+(d+kb)\right)/\left(cx+d\right)^{4}\nonumber \\
= & \frac{P_{4}(x)}{\left(cx+d\right)^{4}}.\label{eq:PzP4}\end{align}
 The solutions are therefore following, using $cr+d=\frac{ad-bc}{\left(a-cz\right)}$
and $dr=\frac{ad-bc}{\left(a-cz\right)^{2}}dz$: \begin{align}
\eta-\eta_{B}= & \int_{R}^{r}\frac{1}{\sqrt{P_{F}(z)}}\frac{1}{\left(cr+d\right)^{2}}dr\nonumber \\
= & \frac{F(\frac{ar+b}{cr+d},k)-F(\frac{aR+b}{cR+d},k)}{\left(ad-bc\right)}.\label{eq:confTSoln}\end{align}
 We then just need to find $a,b,c,d,k$ in terms of $E,M,\Lambda$.
We already have the roots of $P_{4}=P_{3,f}r\frac{\Lambda}{3}$ from
Appendix \ref{sec:Appendix-A:-Roots} and we can write from Eq.~(\ref{eq:PzP4})
\begin{gather}
\begin{array}{rlrl}
r_{1}= & -{\displaystyle \frac{d-b}{c-a}}, & r_{2}= & -{\displaystyle \frac{d+b}{c+a}},\\
r_{3}\vphantom{\frac{\frac{\frac{\frac{}{}}{}}{}}{c}}= & -{\displaystyle \frac{d-kb}{c-ka}}, & r_{4}= & -{\displaystyle \frac{d+kb}{c+ka}}.\end{array}\end{gather}
 We can obtain expressions for $d$ and $b$, isolating them in the
first and second pairs of roots:\begin{gather}
\left\{ \begin{array}{rl}
d= & {\displaystyle -\frac{r_{1}(c-a)+r_{2}(c+a)}{2}}\\
b= & {\displaystyle \frac{r_{1}(c-a)-r_{2}(c+a)}{2}}\end{array}\right.\qquad\qquad\qquad\quad\label{eq:db12}\\
\qquad\qquad\qquad\left.\begin{array}{rl}
= & {\displaystyle -\frac{r_{3}(c-ka)+r_{4}(c+ka)}{2}}\\
= & {\displaystyle \frac{r_{3}(c-ka)-r_{4}(c+ka)}{2k}}\end{array}\right\} .\label{eq:db34}\end{gather}
 Equating the two ways of writing $b+d$, we obtain a linear relation
between $c$ and $a$,\begin{align}
c= & \frac{r_{3}k(1-k)+r_{4}k(1+k)-2kr_{2}}{r_{3}(1-k)+2kr_{2}-r_{4}(1+k)}a.\label{eq:cLina}\end{align}
 Now recall that the factors of $x^{4}$ and $x^{0}$ in $P_{4}$
are, respectively,\begin{gather}
(c^{2}-a^{2})(c^{2}-k^{2}a^{2})=\frac{\Lambda}{3},\label{eq:Coefx4}\\
r_{1}r_{2}r_{3}r_{4}=0.\label{eq:Coefx0}\end{gather}
 The cosmological constant means from Eq.~(\ref{eq:Coefx4}) that
neither $c=\pm a$ nor $c=\pm ka$, while Eq.~(\ref{eq:Coefx0})
entails that one of the roots is 0. If we choose $r_{4}=0$, then
we have $d=-kb$ and therefore, from Eqs.~(\ref{eq:db12}), $d+kb=0$
yields\begin{align}
\frac{c}{a}=\frac{r_{1}(1-k)-r_{2}(1+k)}{r_{1}(1-k)+r_{2}(1+k)},\end{align}
 so with Eq.~(\ref{eq:cLina}) and $r_{4}=0$, we obtain a third
degree polynomial in $k$ (recall $k\ne1$ for nondegeneracy of $P_{F}$)
\begin{gather}
\left(k-1\right)\left\{ \left(k+\frac{2r_{1}r_{2}-r_{1}r_{3}-r_{2}r_{3}}{r_{1}r_{3}-r_{2}r_{3}}\right)^{2}+1\right.\nonumber \\
\left.-\left(\frac{2r_{1}r_{2}-r_{1}r_{3}-r_{2}r_{3}}{r_{1}r_{3}-r_{2}r_{3}}\right)^{2}\right\} =0\end{gather}
\vspace{-0.5cm}
\begin{flalign}
 & \Rightarrow k\nonumber \\
 & =\frac{2r_{1}r_{2}-r_{1}r_{3}-r_{2}r_{3}}{r_{2}r_{3}-r_{1}r_{3}}\pm\sqrt{\left(\frac{2r_{1}r_{2}-r_{1}r_{3}-r_{2}r_{3}}{r_{1}r_{3}-r_{2}r_{3}}\right)^{2}-1}.\end{flalign}
 We also can rewrite the condition (\ref{eq:cLina}) to obtain $a$
with Eq.~(\ref{eq:Coefx4}): the positivity of $\Lambda$ in Eq.~(\ref{eq:Coefx4}),
\begin{align}
\frac{\Lambda}{3}= & \frac{4k^{2}\left(1-k^{2}\right)^{2}\left[\left(1-k\right)^{2}r_{3}+4r_{2}k\right]\left[r_{3}-r_{2}\right]r_{2}r_{3}}{\left[2r_{2}k+\left(1-k\right)r_{3}\right]^{4}}a^{4},\end{align}
 imposes to choose $r_{3}>r_{2}>0$, and thus\begin{align}
a= & \pm\left[2r_{2}k+\left(1-k\right)r_{3}\right]\sqrt{\frac{\sqrt{\frac{\Lambda}{3\left[\left(1-k\right)^{2}r_{3}+4r_{2}k\right]\left[r_{3}-r_{2}\right]r_{2}r_{3}}}}{2k\left|1-k^{2}\right|}}.\label{eq:aSol}\end{align}
 We deduce then $c$ from Eq.~(\ref{eq:cLina})\begin{align}
c= & \pm k\left[\left(1-k\right)r_{3}-2r_{2}\right]\sqrt{\frac{\sqrt{\frac{\Lambda}{3\left[\left(1-k\right)^{2}r_{3}+4r_{2}k\right]\left[r_{3}-r_{2}\right]r_{2}r_{3}}}}{2k\left|1-k^{2}\right|}},\label{eq:cSol}\end{align}
 derive $b$ from including the solutions (\ref{eq:aSol},\ref{eq:cLina})
in its expression in Eq.~(\ref{eq:db12})\begin{align}
b= & \mp\frac{\left[4r_{2}k+\left(1-k\right)^{2}r_{3}\right]r_{1}+\left[\left(1-k^{2}\right)r_{3}\right]r_{2}}{2}\nonumber \\
 & \times\sqrt{\frac{\sqrt{\frac{\Lambda}{3\left[\left(1-k\right)^{2}r_{3}+4r_{2}k\right]\left[r_{3}-r_{2}\right]r_{2}r_{3}}}}{2k\left|1-k^{2}\right|}},\end{align}
 and obtain d with our choice of $r_{4}=0$ that induces $d=-kb$\vspace{-0.01cm}
 \begin{align}
d= & \pm\frac{k\left[4r_{2}k+\left(1-k\right)^{2}r_{3}\right]r_{1}+k\left[\left(1-k^{2}\right)r_{3}\right]r_{2}}{2}\nonumber \\
 & \times\sqrt{\frac{\sqrt{\frac{\Lambda}{3\left[\left(1-k\right)^{2}r_{3}+4r_{2}k\right]\left[r_{3}-r_{2}\right]r_{2}r_{3}}}}{2k\left|1-k^{2}\right|}.}\end{align}
Inputting the values of the roots from Appendix \ref{sec:Appendix-A:-Roots},
and the values of the transformation coefficients $a,\, b,\, c$,
and $d$ into Eq.~(\ref{eq:cosmTSoln}) yields the conformal time
evolution solution that can be related to the cosmic time according
to\begin{align}
t-t_{B}=\int_{\eta_{B}}^{\eta}rd\eta & =\int_{R}^{r}r\frac{\partial}{\partial r}\left(\frac{F(\frac{ar+b}{cr+d},k)}{\left(ad-bc\right)}\right)dr.\label{eq:cosmTSoln}\end{align}
Therefore there is an analytic solution to the $\Lambda$LTB model
(see also Lemaître \citep{Lemaitre33}).



\end{comment}
{}

\bibitem{Sciama:1953zz} D.~W.~Sciama, 
 Mon.\ Not.\ Roy.\ Astron.\ Soc.\textbf{ 113}, 34 (1953). 


\bibitem{Dicke:1961ma} R.~H.~Dicke, 
 Nature (London) \textbf{192}, 440 (1961). 

%
{}

\bibitem{Brans:1961sx} C.~Brans and R.~H.~Dicke, 
 Phys.\ Rev. \textbf{124}, 925 (1961). 

\bibitem{Barrow:1994nx} J.~D.~Barrow and J.~P.~Mimoso, 
 Phys.\ Rev.\ D \textbf{50}, 3746 (1994). 


\bibitem{Mimoso:1994wn} J.~P.~Mimoso and D.~Wands, 
 Phys.\ Rev.\ D \textbf{51}, 477 (1995)
. 


\bibitem{Iguchi:2004yh} H.~Iguchi, T.~Harada and F.~C.~Mena,
 Classical\ Quantum\ Gravity\ \textbf{22}, 841 (2005)
. 

\bibitem{ES} A. Einstein and E. G. Straus, Rev.~Mod.~Phys. \textbf{17},
120 (1945); \textbf{18}, 148 (1946).


\bibitem{Bonnor1974} 
W. B. Bonnor, Mon.\ Not.\ Roy.\ Astron.\ Soc. \textbf{\noun{167}},
55 
 (1974). 
%
{}

\bibitem{Bonnor 1996} 
W. B. Bonnor, Mon.\ Not.\ Roy.\ Astron.\ Soc. \textbf{282}, 1467 
 (1996). 

\bibitem{Fayos} F. Fayos, J. M. M. Senovilla, and R. Torres, Phys.~Rev.~D
\textbf{54}, 4862 (1996).

\bibitem{Krasinski} A.~Krasi\'{n}ski, \emph{Physics in an Inhomogeneous
Universe} (Cambridge University Press, Cambridge 1997).


\bibitem{Krazinski 2001} 
A. Krasi\'{n}ski and C. Hellaby, Phys. Rev. D \textbf{65}, 023501
(2001).


\bibitem{Krasinski:2003yp} A.~Krasinski and C.~Hellaby, 
 Phys.\ Rev.\ D \textbf{69}, 023502 (2004)
. 


\bibitem{Hellaby:2005ut} C.~Hellaby and A.~Krasinski, 
 Phys.\ Rev.\ D \textbf{73}, 023518 (2006)
. 


\bibitem{Mena:2004ck} F.~C.~Mena, B.~C.~Nolan and R.~Tavakol,
 Phys.\ Rev.\ D \textbf{70}, 084030 (2004)
. 

\bibitem{Herrera92}L. Herrera, Phys.\ Lett.\ A \textbf{165}, 206
(1992).

\bibitem{Herrera} A. Abreu, H. Hernandez, and L. A. Nunez, Classical\ Quantum\ Gravity
\textbf{24}, 4631 
 (2007); A. Di Prisco, L. Herrera, and V. Varela, Gen.\ Relativ.\ Gravit.
\textbf{29}, 1239 (1997); L. Herrera and N. O. Santos, Phys.~Rep.
\textbf{286}, 53 (1997).

\bibitem{DiPrisco94}A. Di Prisco, L. Herrera, E. Fuenmayor and V.
Varela, Phys.\ Lett.\ A \textbf{195}, 23 (1994).

%
{}

\bibitem{Adler et al 05} R.~J.~Adler, J.~D.~Bjorkem, P.~Chen,
and J.~S.~Liu, 
arXiv:gr-qc/0502040
.

%
{}

\bibitem{LaskyLun06b}P. D. Lasky and A. W. C. Lun, Phys. Rev. D \textbf{74},
084013 (2006).


\bibitem{LaskyLun06+} P.~D.~Lasky and A.~W.~C.~Lun, 
 Phys.\ Rev.\ D \textbf{75}, 024031 (2007)
; 
 Phys.\ Rev.\ D \textbf{75}, 104010 (2007)
; 
 P.~Lasky and A.~Lun, 
 arXiv:0711.4830
. 

%
{}

\bibitem[m]{MisnerSharp}C. W. Misner and D. H. Sharp, Phys.~Rev.~B
\textbf{136}, 571 (1964).

\bibitem{Delliou:2009dm} M.~Le~Delliou and J.~P.~Mimoso, 
 AIP Conf.\ Proc. \textbf{1122}, 316 (2009)
. 
%
{}


\bibitem{Arnowitt:1962hi} R.~L.~Arnowitt, S.~Deser and C.~W.~Misner,
 arXiv:gr-qc/0405109
; 
 Phys.\ Rev.\ \textbf{117}, 1595 (1960). 


\bibitem{EllisElst98} G.~F.~R.~Ellis and H.~van Elst, 
 NATO Adv.\ Study Inst.\ Ser.\ C.,\ Math.\ Phys.\ Sci. \textbf{541},
1 (1999)
. 

\bibitem{Maartens97}R. Maartens, Phys.~Rev.~D \textbf{55}, 463
(1997).

\bibitem{LeDH03}M. Le Delliou and R. N. Henriksen, Astron.~Astrophys.
\textbf{408}, 27 (2003), and references therein.

\bibitem{NFW}J.~F.~Navarro, C.~S.~Frenk and S.~D.~M.~White,
Astrophys.\ J. \textbf{462}, 563 (1996).

\bibitem[Le Delliou, 2001]{LeDPhD}M. Le Delliou, PhD thesis, Queen's
University, Kingston, Canada, 2001.

%
{}

\bibitem{Lemaitre33}G.~Lemaître, 
de Bruxelles 
, 
Ann.~Soc.~Sci.~B. A\textbf{ 53}, 51 (1933); printed in Gen.\ Relativ.\ Gravit.,
\textbf{29}, 641 (1997).

\bibitem{Matravers:2000cu} D.~R.~Matravers and N.~P.~Humphreys,
 Gen.\ Relativ.\ Gravit., \textbf{33}, 531 (2001)
. 

\end{thebibliography}
{\large 
}{\large \par}

%
{}


\end{document}